\documentclass[%
 reprint,
superscriptaddress,
 amsmath,amssymb,
 aps,
]{revtex4-2}

\usepackage[utf8]{inputenc}
\usepackage{graphicx}
\usepackage{color}
\usepackage{physics}
\usepackage{braket}
\usepackage{comment}
\usepackage{mathrsfs}
\usepackage{enumerate}

\usepackage[version=4]{mhchem}
\usepackage[breaklinks=true]{hyperref}
\usepackage{lipsum}
\usepackage{algorithmic}
\usepackage{algorithm}

\newcommand{\bmvec}[1]{{\mbox{\boldmath $#1$}}}
\newcommand{\mr}[1]{\mathrm{#1}}

\newcommand{\mcl}[1]{\mathcal{#1}}

\newcommand{\bbN}{\mathbb{N}}

\newcommand{\ad}{\mathrm{ad}}

\newcommand{\poly}[1]{\mathrm{poly} \left( #1 \right)}
\newcommand{\polylog}[1]{\mathrm{polylog} \left( #1 \right)}

\date{\today}
\usepackage{amsthm}
\theoremstyle{definition}
\newtheorem{theorem}{Theorem}[]

\newtheorem{proposition}[theorem]{Proposition}

\newtheorem{corollary}[theorem]{Corollary}

\newtheorem*{theorem*}{Theorem}
\newtheorem*{proposition*}{Proposition}

\begin{document}
\title{Trotterization is substantially efficient for low-energy states}

\author{Kaoru Mizuta}
\email{mizuta@qi.t.u-tokyo.ac.jp}
\affiliation{Department of Applied Physics, Graduate School of Engineering, The University of Tokyo, Hongo 7-3-1, Bunkyo, Tokyo 113-8656, Japan}
\affiliation{Photon Science Center, Graduate School of Engineering, The University of Tokyo, Hongo 7-3-1, Bunkyo, Tokyo 113-8656, Japan}
\affiliation{RIKEN Center for Quantum Computing (RQC), Hirosawa 2-1, Wako, Saitama 351-0198, Japan}

\author{Tomotaka Kuwahara}
\affiliation{Analytical Quantum Complexity RIKEN Hakubi Research Team, RIKEN Center for Quantum Computing (RQC), Wako, Saitama 351-0198, Japan }
\affiliation{RIKEN Cluster for Pioneering Research (CPR), Wako, Saitama 351-0198, Japan}
\affiliation{PRESTO, Japan Science and Technology (JST), Kawaguchi, Saitama 332-0012, Japan }

\begin{abstract}
Trotterization is one of the central approaches for simulating quantum many-body dynamics on quantum computers or tensor networks.
In addition to its simple implementation, recent studies have revealed that its error and cost can be reduced if the initial state is closed in the low-energy subspace.
However, the improvement by the low-energy property diminishes rapidly as the Trotter order grows in the previous studies, and thus, it is mysterious whether there exists genuine advantage of low-energy initial states.
In this Letter, we resolve this problem by proving the optimal error bound and cost of Trotterization for low-energy initial states.
For generic local Hamiltonians composed of positive-semidefinite terms, we show that the Trotter error is at most linear in the initial state energy $\Delta$ and polylogarithmic in the system size $N$.
As a result, the computational cost becomes substantially small for low-energy states with $\Delta \in o(Ng)$ compared to the one for arbitrary initial states, where $g$ denotes the energy per site and $Ng$ means the whole-system energy.
Our error bound and cost of Trotterization achieve the theoretically-best scaling in the initial state energy $\Delta$.
In addition, they can be partially extended to weakly-correlated initial states having low-energy expectation values, which are not necessarily closed in the low-energy subspace.
Our results will pave the way for fast and accurate simulation of low-energy states, which are one central targets in condensed matter physics and quantum chemistry.  

\end{abstract}
\maketitle

\textit{Introduction.---} Simulating time-evolution of quantum many-body systems, called Hamiltonian simulation, is one of the central fields in quantum computation, whose applications range from condensed matter physics to quantum chemistry.
When simulating the dynamics of a size-$N$ Hamiltonian over time $t$ within an allowable error $\varepsilon \in (0,1)$, the ultimate goal is to construct quantum algorithms having as few computational resources as possible in $N,t$, and $1/\varepsilon$.
To this goal, various quantum algorithms have been developed in the past few decades, such as Trotterization \cite{Lloyd1996-ko}, linear combination of unitaries (LCU) \cite{Berry-prl2015-LCU}, and quantum singular value transform (QSVT) \cite{Low2019-qubitization,Gilyen2019-qsvt,Martyn2021-grand-unif}.
Trotterization is one of the most famous approaches among them.
It is feasible in near-term quantum computers owing to its simple circuit structures \cite{Maller-prx2016-scalable}. In addition, it is also promising for future large-scale quantum computers since it has better gate complexity in the system size $N$ than other algorithms \cite{childs-prl2019-pf,childs2021-trotter}.

A central open question is whether Trotterization can provably offer a fundamental advantage when the dynamics begin with some interesting classes of initial states.
In general, the gate complexity is determined by the worst-case error among all the possible initial states, such as the commutator-scaling error making  Trotterization efficient in the system size \cite{childs2021-trotter}.
On the other hand, we are often interested in some physically-motivated classes of initial states and this generally overestimates the error and the cost.
For instance, Trotterization with Haar random or highly-entangled initial states hosts better scaling than the one for arbitrary states \cite{Zhao-prl2022-random,zhao2024-entanglement}.
These examples suggest that the structure of the initial state can play a crucial role in improving simulation efficiency. In this context, low-energy initial states are promising candidates, as they naturally arise in low-temperature phenomena in physics and chemistry~\cite{Low2017-ig-low-energy,Gu_2021-low-energy,Zlokapa_2024-low-energy,Sahinoglu2021-ng-low-energy,Gong2024-wr-low-energy,Hejazi2024-xs-low-energy,hejazi-2024arXiv-low-energy-qpe}.
Recent studies \cite{Sahinoglu2021-ng-low-energy,Gong2024-wr-low-energy,Hejazi2024-xs-low-energy} have revealed that Trotterization can have the reduction in its error and cost for low-energy initial states.
They show that the gate complexity is improved in the system size $N$ when the initial state energy $\Delta$ is smaller than $\order{N^{1/(p+1)}g}$, where $p$ denotes the Trotter order and $g$ means the maximum energy per site.
However, their results are quite restricted: the advantageous low-energy regime $\Delta \in o(N^{1/(p+1)}g)$ and the improvement in the system size shrink as the Trotter order $p$ grows.
On the contrary, their estimates give worse errors and cost for low-energy initial states with $N^{1/(p+1)}g \lesssim \Delta < Ng$ than the one for arbitrary initial states.
It is mysterious whether there exists a provable advantage in generic low-energy initial states whose energy $\Delta$ is smaller than the whole-system energy $Ng$ (i.e., $\Delta \in o(Ng)$).

In this Letter, we provide the optimal solution for Trotterization for generic low-energy initial states with $\Delta \in o(Ng)$, rigorously establishing that they offer a substantial and intrinsic advantage in Hamiltonian simulation (See Table \ref{Table:Comparison}).
We mainly focus on local Hamiltonians composed of positive-semidefinite finite-range interactions and consider initial states closed in the low-energy subspace as well as the previous studies.
We first derive the optimal error bound for low-energy initial states, which accurately reflecting the commutator scaling in Trotterization.
Compared to the previous studies \cite{Sahinoglu2021-ng-low-energy,Gong2024-wr-low-energy,Hejazi2024-xs-low-energy}, our error bound is linear in the initial state energy $\Delta$ and achieves exponentially better scaling explicitly in the system size $N$.
While the Trotter number for arbitrary initial states amounts to $r \in \order{gt (Ngt/\varepsilon)^{1/p}}$ \cite{childs2021-trotter}, we obtain
\begin{equation}\label{Eq:Trotter_number_ours}
    r \sim gt \left( \frac{\Delta t + gt \log (N/\varepsilon)}{\varepsilon}\right)^{\frac1p},
\end{equation}
as the one for low-energy initial states.
Cost reduction appears in the broadest low-energy regime $\Delta \in o(Ng)$ regardless of the Trotter order $p$.
At the same time, initial states with extremely low energy $\Delta \in \polylog{N}g$ require only $\polylog{N}$ gate complexity under the fixed time $t$ and error $\varepsilon$, implying the exponential speedup for the sufficiently low-energy regime.
We emphasize that our error bound and cost respectively achieve the theoretically-best scaling in the initial state energy $\Delta$ since they can reproduce the optimal ones of Trotterization for arbitrary initial states \cite{zhao2024-entanglement} with $\Delta=Ng$.

Our results can be extended to various cases:
The improved Trotter number is obtained for generic local Hamiltonians involving long-range interactions, where Eq. (\ref{Eq:Trotter_number_ours}) acquires at most polylogarithmic corrections.
We also show that the cost reduction is present for weakly-correlated states with a low-energy expectation value, $\braket{\psi|H|\psi} \in o(Ng)$ as a corollary of the main results.
Our results reveal the substantial advantage of low-energy states in quantum computation, which is of central interest in physics and chemistry.
These findings underscore the broader principle that the structure of the initial state can crucially influence simulation efficiency.
More broadly, our work initiates a framework for initial-state–dependent Hamiltonian simulation, where algorithmic performance is tailored to physically meaningful input classes.
Our result provides the first optimal bound in this direction for low-energy states, and the commutator-scaling technique developed here is flexible enough to extend to other families such as weakly-correlated thermal or highly-entangled states \cite{Zhao-prl2022-random,zhao2024-entanglement}.

\textit{Setup and notations.---}
We first specify the setup and some notations.
Throughout this Letter, we focus on a $N$-qubit Hamiltonian $H$ on a lattice $\Lambda = \{ 1,2,\cdots, N \}$.
We assume that $H$ is a $k$-local Hamiltonian with $k \in \order{1}$ written by
\begin{equation}\label{Eq:Hamiltonian}
    H = \sum_{X \subset \Lambda; |X| \leq k} h_X, \quad h_X \geq 0
\end{equation}
where each $h_X$ nontrivially acts on a domain $X$.
The positive-semidefiniteness of $h_X$ can be always satisfied by the shift $h_X \to h_X + \norm{h_X}$.
We define the extensiveness $g$ by a value such that
\begin{equation}\label{Eq:def_extensiveness}
    \sum_{X \subset \Lambda; X \ni i} \norm{h_X} \leq g, \quad ^\forall i \in \Lambda,
\end{equation}
which means the maximal energy scale per site.
The scaling of $g$ depends on the range of interactions.
When a Hamiltonian has finite-range interactions (i.e., each $X$ giving $h_X \neq 0$ includes sites within some constant distance), we have the extensiveness $g \in \order{N^0}$.
When a Hamiltonian has long-range interactions (i.e., every interactions between two sites $i,j$, given by $\sum_{X \ni i,j} \norm{h_X}$ decays as their distance $\order{\mr{dist}(i,j)^{-\nu}}$), the extensiveness scales as
\begin{equation}\label{Eq:Extensiveness}
    g \in \begin{cases}
        \order{N^0} & (\text{if long-ranged, $\nu >d$}) \\
        \order{\log N} & (\text{if long-ranged, $\nu =d$}) \\
        \order{N^{1-\frac\nu{d}}} & (\text{if long-ranged, $\nu<d$}),
    \end{cases}
\end{equation}
where $d$ denotes the spatial dimension of the lattice $\Lambda$.

Next, we introduce Trotterization as an approach to Hamiltonian simulation.
We begin with decomposing the Hamiltonian $H$ into $\Gamma$ terms by
\begin{equation}\label{Eq:trotter_partition}
    H = \sum_{\gamma=1}^\Gamma H_\gamma, \quad H_\gamma = \sum_{X \subset \Lambda} h_X^\gamma.
\end{equation}
Each pair in $\{ h_X^\gamma \geq 0 \}_X$ commutes with each other, and each term $h_X$ is reproduced by $h_X = \sum_\gamma h_X^\gamma$.
The number of partition $\Gamma$ will be a significant factor in our result.
We have $\Gamma \in \order{1}$ for generic local Hamiltonians with finite-range interactions, and some Hamiltonians with long-range interactions (See Supplementary Materials \ref{SecS:num_terms} in detail \cite{Supple}).
The $p$-th order Trotter formula $T_p(t)$ is defined by the product of $c_p \Gamma$ time-evolution operators satisfying the order condition as follows,
\begin{equation}\label{Eq:product_formula}
    T_p(t) = \prod_{v=1}^{c_p \Gamma} e^{-iH_{\gamma_v} \alpha_v t} = e^{-iHt} + \order{t^{p+1}},
\end{equation}
We assume that the real numbers $\{ \alpha_v \}_v$ satisfy $|\alpha_v| \leq 1$ without loss of generality.
We have $c_p \in \order{1}$ as long as $p \in \order{1}$.
The Lie-Suzuki-Trotter formula is one of the examples, whose first and second orders are given by
\begin{equation}\label{Eq:Lie-Suzuki-Trotter}
    T_1(t) = e^{-iH_\Gamma t} \cdots e^{-iH_1 t}, \quad T_2(t) = T_1(-t/2)^\dagger T_1(t/2).
\end{equation}  
The number $c_p$ is given by $c_1=1$ and $c_p=2 \cdot 5^{p/2-1}$ for $p \in 2\bbN$ in this case.

For simulating dynamics over large time $t$, we implement Trotterization $T_p(t/r)$ for $r$ times.
The query complexity $r$, called the Trotter number, is determined so that we can achieve the desirable error $\varepsilon$ by
\begin{equation}
    \norm{e^{-iHt}-\{T_p(t/r)\}^r} \leq r \norm{e^{-iHt/r}-T_p(t/r)} \leq \varepsilon.
\end{equation}
The error bound of Trotterization, $\norm{e^{-iHt/r}-T_p(t/r)}$, has been recently revealed in a unified way \cite{childs2021-trotter}.
Owing to its commutator scaling, the error and the cost of Trotterization have better scaling in the system size $N$, as shown in the first row of Table \ref{Table:Comparison}.

\begin{table*}[]
    \centering
    \begin{tabular}{c|c|c|c|}
         Initial state & Error bound & Modified energy bound $\Delta'$ & Trotter number $r$ \\ \hline
        \begin{tabular}{c} Arbitrary state \\ (Ref. \cite{childs2021-trotter})\end{tabular} & \( \displaystyle (gt)^p Ngt \)& --- & \( \displaystyle gt \left( \frac{Ngt}{\varepsilon}\right)^{\frac1p} \) \\ \hline

        \begin{tabular}{c} Low-energy state \\
        (Ref. \cite{Sahinoglu2021-ng-low-energy,Gong2024-wr-low-energy}) \end{tabular} & \( \displaystyle (\Delta't)^{p+1}+\epsilon \) & \( \displaystyle \Delta + \order{g \log (N/\epsilon) + Ng (gt)} \) & \( \displaystyle
            \Delta t \left(\frac{\Delta t}{\varepsilon}\right)^{\frac1p}+ gt \left(\frac{gt}{\varepsilon}\right)^{\frac1{2p+1}} N^{\frac12+\frac1{4p+2}}
        \) \\ \hline

        \begin{tabular}{c} Low-energy state \\
        (Ref. \cite{Hejazi2024-xs-low-energy}) \end{tabular} & \( \displaystyle (\Delta't)^{p+1}+\epsilon \) & \( \displaystyle \Delta + \order{g \log (N/\epsilon) + Ng (gt)^{p+1}} \) & \( \displaystyle
            \Delta t \left(\frac{\Delta t}{\varepsilon}\right)^{\frac1p}+ gt \left( \frac{N^{p+1}gt}{\varepsilon}\right)^{\frac1{(p+1)^2+p}}
         \)\ \\ \hline
            \begin{tabular}{c}
            Low-energy state  \\
            $\Gamma \in \order{1}$ (This work)
         \end{tabular} & \( \displaystyle  (gt)^p \Delta't+\epsilon \) & \( \displaystyle \Delta + \order{g \log (N/\epsilon)} \) & \( \displaystyle
            gt \left(\frac{\Delta t + gt \log (N/\varepsilon)}{\varepsilon} \right)^{\frac1p}
         \)\ \\ \hline
         \begin{tabular}{c} Low-energy state \\
        $\Gamma \notin \order{1}$ (This work) \end{tabular} & \( \displaystyle  \{gt \log (N/\epsilon)\}^p \Delta't+\epsilon \) & \( \displaystyle \Delta + \order{g \log (N/\epsilon)} \) & \( \displaystyle
            gt \left(\frac{\Delta t + gt \log (N/\varepsilon)}{\varepsilon} \right)^{\frac1p} \log (N/\varepsilon)
         \)\ \\ \hline

    \end{tabular} 
    \caption{Error bounds and Trotter numbers for arbitrary or low-energy initial states with the energy smaller than $\Delta$. The value $g$ means the extensiveness given by Eq. (\ref{Eq:def_extensiveness}). The case with $\Gamma \in \order{1}$ includes generic Hamiltonians with finite-range interactions and some Hamiltonians with long-range interactions. The case with $\Gamma \notin \order{1}$ includes the other generic local Hamiltonians.
    }
    \label{Table:Comparison}
\end{table*}

\textit{Trotter error in the low-energy subspace.---}
We hereby provide the first main result on the suppressed Trotter errors for low-energy initial states, which will lead to the better cost of Hamiltonian simulation from the low-energy subspace.

For a given Hamiltonian $H$, we define a projection to a subspace with the energy lower than $\Delta$ by
\begin{equation}\label{Eq:proj_low_energy}
    \Pi_{\leq \Delta} = \sum_{n; E_n \leq \Delta} \ket{E_n}\bra{E_n}.
\end{equation}
$(E_n,\ket{E_n})$ gives the spectral decomposition of $H$ by $H= \sum_n E_n \ket{E_n}\bra{E_n}$.
Then, we suppose that the initial state $\ket{\psi}$ is closed in this subspace as $\Pi_{\leq \Delta} \ket{\psi} = \ket{\psi}$.
We mean low-energy by the assumption that $\Delta$ is sufficiently small compared to $Ng$, which gives the whole energy scale by
\begin{equation}
    \norm{H} \leq \sum_{i \in \Lambda} \sum_{X; X \ni i} \norm{h_X} \leq Ng.
\end{equation}
We focus on the worst case error of Trotterization for all the possible low-energy initial states, defined as follows,
\begin{eqnarray}
    \varepsilon_{p,\Delta} (t) &\equiv& \max_{\substack{\ket{\psi}; \norm{\ket{\psi}}=1, \\  \Pi_{\leq \Delta}\ket{\psi}=\ket{\psi}}} \left( \norm{(e^{-iHt}-T_p(t)) \ket{\psi}}\right)\\
    &=& \norm{(e^{-iHt} -T_p(t))\Pi_{\leq \Delta}}.
\end{eqnarray}
Our first main result is the upper bound on $\varepsilon_{p,\Delta} (t)$, which is better than any previous studies \cite{Sahinoglu2021-ng-low-energy,Gong2024-wr-low-energy,Hejazi2024-xs-low-energy} as far as we know.

\begin{theorem}\label{Thm:Error_low_energy}
\textbf{}

Let $\epsilon$ be an arbitrary value in $(0,1)$.
Then, there exists a value,
\begin{equation}\label{Eq:Modified_energy_bound}
    \Delta' = \Delta + \order{g \log (N/\epsilon)},
\end{equation}
such that the Trotter error from low-energy initial states is bounded by
\begin{equation}\label{Eq:Error_bound_low_energy}
    \varepsilon_{p,\Delta}(t) \in \begin{cases}
        \order{(gt)^p \Delta't + \epsilon} & (\text{if $\Gamma \in \order{1}$}) \\
        \order{\{gt \log (N/\epsilon)\}^p \Delta't + \epsilon} & (\text{otherwise})
    \end{cases}
\end{equation}
for small time $t$ satisfying
\begin{equation}\label{Eq:time_condition}
    |t| \in \begin{cases}
        \order{g^{-1}} & (\text{if $\Gamma \in \order{1}$}) \\
        \order{\{g\log(N/\epsilon)\}^{-1}} & (\text{otherwise}).
    \end{cases} 
\end{equation}
\end{theorem}

We note that Eqs. (\ref{Eq:Modified_energy_bound}), (\ref{Eq:Error_bound_low_energy}), and (\ref{Eq:time_condition}) include constant factors dependent on the locality $k \in \order{1}$ and the order $p \in \order{1}$.
See Supplementary Materials \ref{SecS:trotter_error} for the detailed proof and the exact bound \cite{Supple}.
Here, we discuss the meaning of Theorem \ref{Thm:Error_low_energy}.
Our result depends on the number of the partition $\Gamma$ in Eq. (\ref{Eq:trotter_partition}).
We focus on the first case, $\Gamma \in \order{1}$, which applies to generic local Hamiltonians with finite-range interactions.
When the initial state has no energy restriction, the worst-case Trotter error increases as the whole energy scale $Ng$ as the first row of Table \ref{Table:Comparison}.
However, when the initial state is tied to the low-energy subspace, such large energy scale seems not to be relevant.
Theorem \ref{Thm:Error_low_energy} rigorously validates this expectation, where the energy scale $Ng$ is replaced by $\Delta'$, a value close to $\Delta$.
The deviation $\Delta'-\Delta \in \order{g\log(N/\epsilon)}$ comes from the leakage from the low-energy subspace since the Trotterization $T_p(t)$ does not exactly preserve $H$.
The Trotter error is dominated by the initial state energy $\Delta$ rather than the system size $N$, which is also numerically demonstrated in some local Hamiltonians (See Supplementary Materials \ref{SecS:Numerical} \cite{Supple}).

We also make comparison with the previous results in the second and third rows \cite{Sahinoglu2021-ng-low-energy,Hejazi2024-xs-low-energy}, briefly mentioning about our derivation.
Our central contributions compared to them are the reflection of the commutator scaling in the low-energy space and the evaluation of the leakage caused by the Trotter error.
First, the commutator scaling is the characteristics of the Trotter error \cite{childs2021-trotter}, which states that the error bound for low-energy initial states is dominated by
\begin{equation}\label{Eq:com_low_energy}
    \sum_{\gamma_0,\cdots,\gamma_q} \norm{\Pi_{\leq \Delta'}[H_{\gamma_q},\cdots,[H_{\gamma_1},H_{\gamma_0}]] \Pi_{\leq \Delta'}} t^{q+1},
\end{equation}
for the orders $q \geq p$.
While the nested commutators among local terms $\{ h_X^\gamma \}$ having no overlap with each other vanish, the previous studies fail to catch this feature, which leads to the $1$-norm error scaling $\order{(\Delta't)^{p+1}}$.
By contrast, we successfully prove that Eq. (\ref{Eq:com_low_energy}) scales as $\order{(gt)^q q! \Delta't}$ (See Supplementary Materials \ref{SecS:nested_com} \cite{Supple}).
This reflects the property of the nested commutators as well as the case with no energy restriction, $\order{(gt)^q q! Ngt}$ \cite{childs2021-trotter}.

The second difference is the leakage from the low-energy subspace, related to the choice of $\Delta'$ by Eq. (\ref{Eq:Modified_energy_bound}).
Both of the previous and our studies rely on the following inequality for an operator $O_X$ on a domain $X \subset \Lambda$ \cite{Arad2016-ak-excitation}, 
\begin{equation}\label{Eq:Suppressed_excitation}
    \norm{\Pi_{>\Delta'} O_X \Pi_{\leq \Delta}} \leq \norm{O_X} e^{-\frac{\Delta'-\Delta-3g|X|}{4kg}}.
\end{equation}
It states that the excitation by a local operator $O_X$ is exponentially suppressed under a local Hamiltonian.
The value $\Delta'$ was determined so that the leakage $\norm{\Pi_{>\Delta'} e^{-iH_{\gamma_v}\alpha_vt}\Pi_{\leq \Delta}}$ \cite{Sahinoglu2021-ng-low-energy,Gong2024-wr-low-energy} or $\norm{\Pi_{>\Delta'}T_p(t) \Pi_{\leq \Delta}}$ \cite{Hejazi2024-xs-low-energy} can be suppressed up to a small value $\epsilon$, where the projection $\Pi_{>\Delta'}$ denotes $1-\Pi_{\leq \Delta'}$.
However, the non-locality of Trotterization results in the overestimation of the leakage $\Delta'-\Delta$, which becomes proportional to $N$.
By contrast, we clarify that the leakage relevant to the Trotter error comes from the nested commutators, expressed by
\begin{eqnarray}
    && \sum_{\gamma_0,\cdots,\gamma_q} \norm{\Pi_{> \Delta'}[H_{\gamma_q},\cdots,[H_{\gamma_1},H_{\gamma_0}]] \Pi_{\leq \Delta}} t^{q+1} \nonumber \\
    && \qquad \qquad \in \order{ (gt)^{q+1} q! N \exp \left( -\frac{\Delta'-\Delta}{4kg}\right)},
\end{eqnarray}
for the orders $q \geq p$.
Owing to the locality of nested commutators, the leakage is exponentially improved in $N$ as Eq. (\ref{Eq:Modified_energy_bound}).

Finally, we emphasize that our error bound proportional to the initial state energy $\Delta$ is optimal.
Recently, it has been shown that there exists a set of a local Hamiltonian and an initial state, whose Trotter error is proportional to $N$ \cite{zhao2024-entanglement}.
This means that the Trotter error for generic local Hamiltonians and arbitrary initial states should be at least proportional to $N$.
Since the error bound for low-energy should cover the one for arbitrary initial states with $\Delta = Ng$, the linear scaling in the initial state energy $\Delta$ is theoretically best.   

\textit{Trotter number for low-energy states.---}
Next, we discuss the cost of Hamiltonian simulation for low-energy initial states.
For simulating quantum dynamics for large time $t$, we repeat the Trotter step $T_p(t/r)$ with the Trotter number $r$.
It is determined so that the accumulated error becomes smaller than the allowable error by
\begin{equation}
    \norm{(e^{-iHt} - \{T_p (t/r)\}^r) \Pi_{\leq \Delta}} \leq r \varepsilon_{p,\Delta}(t/r) \leq \varepsilon,
\end{equation}
where the first inequality comes from $[e^{-iHt}, \Pi_{\leq \Delta}]=0$.
Using Theorem \ref{Thm:Error_low_energy} with setting $\epsilon = \varepsilon/(2r)$, it is sufficient to find $r$ such that
\begin{equation}
    \frac{(gt)^p (\Delta t + gt \log (N r/\varepsilon))}{r^p} \in \order{\varepsilon}
\end{equation}
in the case of $\Gamma \in \order{1}$.
Similar discussion goes for $\Gamma \notin \order{1}$.
Under the natural assumption $t, 1/\varepsilon \in \poly{N}$, we obtain the Trotter number $r$ satisfying
\begin{equation}\label{Eq:trotter_number}
    r \in \begin{cases}
        \order{gt \left( \frac{\Delta t + gt \log (N/\varepsilon)}{\varepsilon} \right)^\frac1p } & (\text{if $\Gamma \in \order{1}$}) \\
        \order{gt \log (N/\epsilon) \left( \frac{\Delta t + gt \log (N/\varepsilon)}{\varepsilon} \right)^\frac1p} & (\text{otherwise}),
    \end{cases}
\end{equation}
whose detailed derivation is discussed in Supplementary Materials \ref{SecS:trotter_number} \cite{Supple}.

We compare our results with the cost for arbitrary initial states.
In the case of $\Gamma \in \order{1}$, Table \ref{Table:Comparison} says that there exists a low-energy regime $\Delta \leq \mr{Const.} \times Ng$, where the gate count estimate for low-energy initial states becomes smaller than the one without energy restriction by a constant factor.
The Trotter number has better scaling for $\Delta \in o(Ng)$, and notably, it can be at most $\polylog{N}$ if there exists a sufficiently low-energy regime $\Delta \in \polylog{N}g$.
Such improvement is present also in the case of $\Gamma \notin \order{1}$ as long as the initial state is restricted to the low-energy $\Delta \in \order{Ng/(\log N)^p}$.
Namely, the Trotterization can be efficient for low-energy states also in generic long-range interacting systems.

Compared to the previous results \cite{Sahinoglu2021-ng-low-energy,Hejazi2024-xs-low-energy}, our results give significant improvement both in $\Delta$ and $N$.
First, the polynomially better scaling in $\Delta$ than the second and third rows in Table \ref{Table:Comparison} comes from our discovery, the presence of the commutator-scaling for low-energy initial states as Eq. (\ref{Eq:com_low_energy}).
Although the previous results tell us the advantage of low-energy states for $\Delta \in \order{N^{1/(p+1)}g}$, which are almost absent for the large order $p$, we clarify the advantage in much broader regimes $\Delta \in \order{Ng}$ [for $\Gamma \in \order{1}$] or $\Delta \in \order{Ng/(\log N)^p}$ [for $\Gamma \notin \order{1}$].
Furthermore, the explicit scaling in the system size $N$ is exponentially improved.
It originates from the accurate evaluation of the leakage $\Delta'-\Delta$ relevant for the Trotter error by Eq. (\ref{Eq:Modified_energy_bound}).
This improvement is essential for concluding the non-vanishing advantage in the system size scaling for the large order $p$.
Reflecting the optimality of our error bound in $\Delta$, the Trotter number by Eq. (\ref{Eq:trotter_number}) is also optimal in $\Delta$.
In addition, there is room for improving its system-size dependence only logarithmically.
Thus, our results settle how Trotterization can be efficient for low-energy initial states, which has been vigorously explored in a series of recent studies.

The previous studies and the discussion so far have focused on initial states completely closed in the low-energy subspace.
However, it is often difficult to examine the initial state satisfies such a low-energy condition.
We partially extend our results also for initial states with a low-energy expectation value $\braket{\psi|H|\psi}$, which is easy to be checked by classical or quantum computation.
As a corollary of our results with the so-called concentration bound \cite{Anshu2016-gc-concentration,Kuwahara2016-gh-Chernoff}, we show that the error and cost of Trotterization reduce when the initial state is weakly correlated and has an energy expectation value $\braket{\psi|H|\psi} \in o(Ng)$.
See Supplementary Materials \ref{SecS:weakly_correlated} \cite{Supple} for its detailed discussion.

\textit{Conclusion and Discussion.---}
In this Letter, we show the substantial efficiency of Trotterization for low-energy initial states.
The error bound of Trotterization is at most proportional to the initial state energy $\Delta$ and explicitly logarithmic in the system size $N$, which indicates that the Trotter number can reduce for initial states only with slight low-energy condition $\Delta \in o(Ng)$.
Our results provide the theoretically-best solution to the error bound and cost optimal in the initial state energy $\Delta$, settling a recent series of studies on Trotterization for low-energy states \cite{Sahinoglu2021-ng-low-energy,Hejazi2024-xs-low-energy}.

As a future direction, our results may be extended to broad variants of Trotterization, such as randomly-compiled Trotterization (called qDRIFT \cite{Campbell-prl2019-qdrift,Gong2024-wr-low-energy}) and multi-product formulas \cite{low2019-mpf,zhuk2024-mpf,aftab2024-mpf,Watson2024-mpf}.
The extension to time-dependent Trotterization \cite{Suzuki1993-general,Wiebe2010-mu,childs-prl2019-pf,mizuta2024arxiv-time-dep-PF,cao2024arxiv-time-dep-PF} is of importance, which will lead to the acceleration of adiabatic state preparation \cite{Aspuru-Guzik2005-adiabatic,Du-prl2010-adiabatic}.
We finally note that, while our results achieve optimal scaling, the initial state energy $\Delta$ is defined upon the shift of each term $h_X \to h_X + \norm{h_X}$ for ensuring its positive-semidefiniteness in general.
It is also significant to seek for possible advantages of low-energy initial states whose energy is measured based on the ground state energy.

\section*{Acknowledgment}

K. M. is supported by JST PRESTO Grant No. JPMJPR235A, JSPS KAKENHI Grant No. JP24K16974, and JST Moonshot R\&D Grant No. JPMJMS2061.
T. K. acknowledges the Hakubi projects of RIKEN. 
T. K. is supported by JST PRESTO (Grant No. JPMJPR2116), Exploratory Research for Advanced Technology (Grant No. JPMJER2302), and JSPS Grants-in-Aid for Scientific Research (No. JP24H00071), Japan.

\bibliography{bibliography.bib}

\begin{thebibliography}{37}%
\makeatletter
\providecommand \@ifxundefined [1]{%
 \@ifx{#1\undefined}
}%
\providecommand \@ifnum [1]{%
 \ifnum #1\expandafter \@firstoftwo
 \else \expandafter \@secondoftwo
 \fi
}%
\providecommand \@ifx [1]{%
 \ifx #1\expandafter \@firstoftwo
 \else \expandafter \@secondoftwo
 \fi
}%
\providecommand \natexlab [1]{#1}%
\providecommand \enquote  [1]{``#1''}%
\providecommand \bibnamefont  [1]{#1}%
\providecommand \bibfnamefont [1]{#1}%
\providecommand \citenamefont [1]{#1}%
\providecommand \href@noop [0]{\@secondoftwo}%
\providecommand \href [0]{\begingroup \@sanitize@url \@href}%
\providecommand \@href[1]{\@@startlink{#1}\@@href}%
\providecommand \@@href[1]{\endgroup#1\@@endlink}%
\providecommand \@sanitize@url [0]{\catcode `\\12\catcode `\$12\catcode `\&12\catcode `\#12\catcode `\^12\catcode `\_12\catcode `\%12\relax}%
\providecommand \@@startlink[1]{}%
\providecommand \@@endlink[0]{}%
\providecommand \url  [0]{\begingroup\@sanitize@url \@url }%
\providecommand \@url [1]{\endgroup\@href {#1}{\urlprefix }}%
\providecommand \urlprefix  [0]{URL }%
\providecommand \Eprint [0]{\href }%
\providecommand \doibase [0]{https://doi.org/}%
\providecommand \selectlanguage [0]{\@gobble}%
\providecommand \bibinfo  [0]{\@secondoftwo}%
\providecommand \bibfield  [0]{\@secondoftwo}%
\providecommand \translation [1]{[#1]}%
\providecommand \BibitemOpen [0]{}%
\providecommand \bibitemStop [0]{}%
\providecommand \bibitemNoStop [0]{.\EOS\space}%
\providecommand \EOS [0]{\spacefactor3000\relax}%
\providecommand \BibitemShut  [1]{\csname bibitem#1\endcsname}%
\let\auto@bib@innerbib\@empty
\bibitem [{\citenamefont {Lloyd}(1996)}]{Lloyd1996-ko}%
  \BibitemOpen
  \bibfield  {author} {\bibinfo {author} {\bibfnamefont {S.}~\bibnamefont {Lloyd}},\ }\bibfield  {title} {\bibinfo {title} {{Universal Quantum Simulators}},\ }\href {https://www.science.org/doi/10.1126/science.273.5278.1073} {\bibfield  {journal} {\bibinfo  {journal} {Science}\ }\textbf {\bibinfo {volume} {273}},\ \bibinfo {pages} {1073} (\bibinfo {year} {1996})}\BibitemShut {NoStop}%
\bibitem [{\citenamefont {Berry}\ \emph {et~al.}(2015)\citenamefont {Berry}, \citenamefont {Childs}, \citenamefont {Cleve}, \citenamefont {Kothari},\ and\ \citenamefont {Somma}}]{Berry-prl2015-LCU}%
  \BibitemOpen
  \bibfield  {author} {\bibinfo {author} {\bibfnamefont {D.~W.}\ \bibnamefont {Berry}}, \bibinfo {author} {\bibfnamefont {A.~M.}\ \bibnamefont {Childs}}, \bibinfo {author} {\bibfnamefont {R.}~\bibnamefont {Cleve}}, \bibinfo {author} {\bibfnamefont {R.}~\bibnamefont {Kothari}},\ and\ \bibinfo {author} {\bibfnamefont {R.~D.}\ \bibnamefont {Somma}},\ }\bibfield  {title} {\bibinfo {title} {Simulating hamiltonian dynamics with a truncated taylor series},\ }\href {https://doi.org/10.1103/PhysRevLett.114.090502} {\bibfield  {journal} {\bibinfo  {journal} {Phys. Rev. Lett.}\ }\textbf {\bibinfo {volume} {114}},\ \bibinfo {pages} {090502} (\bibinfo {year} {2015})}\BibitemShut {NoStop}%
\bibitem [{\citenamefont {Low}\ and\ \citenamefont {Chuang}(2019)}]{Low2019-qubitization}%
  \BibitemOpen
  \bibfield  {author} {\bibinfo {author} {\bibfnamefont {G.~H.}\ \bibnamefont {Low}}\ and\ \bibinfo {author} {\bibfnamefont {I.~L.}\ \bibnamefont {Chuang}},\ }\bibfield  {title} {\bibinfo {title} {Hamiltonian simulation by qubitization},\ }\href {https://quantum-journal.org/papers/q-2019-07-12-163/} {\bibfield  {journal} {\bibinfo  {journal} {Quantum}\ }\textbf {\bibinfo {volume} {3}},\ \bibinfo {pages} {163} (\bibinfo {year} {2019})}\BibitemShut {NoStop}%
\bibitem [{\citenamefont {Gily{\'e}n}\ \emph {et~al.}(2019)\citenamefont {Gily{\'e}n}, \citenamefont {Su}, \citenamefont {Low},\ and\ \citenamefont {Wiebe}}]{Gilyen2019-qsvt}%
  \BibitemOpen
  \bibfield  {author} {\bibinfo {author} {\bibfnamefont {A.}~\bibnamefont {Gily{\'e}n}}, \bibinfo {author} {\bibfnamefont {Y.}~\bibnamefont {Su}}, \bibinfo {author} {\bibfnamefont {G.~H.}\ \bibnamefont {Low}},\ and\ \bibinfo {author} {\bibfnamefont {N.}~\bibnamefont {Wiebe}},\ }\bibfield  {title} {\bibinfo {title} {Quantum singular value transformation and beyond: exponential improvements for quantum matrix arithmetics},\ }in\ \href {https://dl.acm.org/doi/10.1145/3313276.3316366} {\emph {\bibinfo {booktitle} {Proceedings of the 51st Annual {ACM} {SIGACT} Symposium on Theory of Computing}}},\ \bibinfo {series and number} {STOC 2019}\ (\bibinfo  {publisher} {Association for Computing Machinery},\ \bibinfo {address} {New York, NY, USA},\ \bibinfo {year} {2019})\ pp.\ \bibinfo {pages} {193--204}\BibitemShut {NoStop}%
\bibitem [{\citenamefont {Martyn}\ \emph {et~al.}(2021)\citenamefont {Martyn}, \citenamefont {Rossi}, \citenamefont {Tan},\ and\ \citenamefont {Chuang}}]{Martyn2021-grand-unif}%
  \BibitemOpen
  \bibfield  {author} {\bibinfo {author} {\bibfnamefont {J.~M.}\ \bibnamefont {Martyn}}, \bibinfo {author} {\bibfnamefont {Z.~M.}\ \bibnamefont {Rossi}}, \bibinfo {author} {\bibfnamefont {A.~K.}\ \bibnamefont {Tan}},\ and\ \bibinfo {author} {\bibfnamefont {I.~L.}\ \bibnamefont {Chuang}},\ }\bibfield  {title} {\bibinfo {title} {{Grand Unification of Quantum Algorithms}},\ }\href {https://journals.aps.org/prxquantum/abstract/10.1103/PRXQuantum.2.040203} {\bibfield  {journal} {\bibinfo  {journal} {PRX Quantum}\ }\textbf {\bibinfo {volume} {2}},\ \bibinfo {pages} {040203} (\bibinfo {year} {2021})}\BibitemShut {NoStop}%
\bibitem [{\citenamefont {O'Malley}\ \emph {et~al.}(2016)\citenamefont {O'Malley}, \citenamefont {Babbush}, \citenamefont {Kivlichan}, \citenamefont {Romero}, \citenamefont {McClean}, \citenamefont {Barends}, \citenamefont {Kelly}, \citenamefont {Roushan}, \citenamefont {Tranter}, \citenamefont {Ding}, \citenamefont {Campbell}, \citenamefont {Chen}, \citenamefont {Chen}, \citenamefont {Chiaro}, \citenamefont {Dunsworth}, \citenamefont {Fowler}, \citenamefont {Jeffrey}, \citenamefont {Lucero}, \citenamefont {Megrant}, \citenamefont {Mutus}, \citenamefont {Neeley}, \citenamefont {Neill}, \citenamefont {Quintana}, \citenamefont {Sank}, \citenamefont {Vainsencher}, \citenamefont {Wenner}, \citenamefont {White}, \citenamefont {Coveney}, \citenamefont {Love}, \citenamefont {Neven}, \citenamefont {Aspuru-Guzik},\ and\ \citenamefont {Martinis}}]{Maller-prx2016-scalable}%
  \BibitemOpen
  \bibfield  {author} {\bibinfo {author} {\bibfnamefont {P.~J.~J.}\ \bibnamefont {O'Malley}}, \bibinfo {author} {\bibfnamefont {R.}~\bibnamefont {Babbush}}, \bibinfo {author} {\bibfnamefont {I.~D.}\ \bibnamefont {Kivlichan}}, \bibinfo {author} {\bibfnamefont {J.}~\bibnamefont {Romero}}, \bibinfo {author} {\bibfnamefont {J.~R.}\ \bibnamefont {McClean}}, \bibinfo {author} {\bibfnamefont {R.}~\bibnamefont {Barends}}, \bibinfo {author} {\bibfnamefont {J.}~\bibnamefont {Kelly}}, \bibinfo {author} {\bibfnamefont {P.}~\bibnamefont {Roushan}}, \bibinfo {author} {\bibfnamefont {A.}~\bibnamefont {Tranter}}, \bibinfo {author} {\bibfnamefont {N.}~\bibnamefont {Ding}}, \bibinfo {author} {\bibfnamefont {B.}~\bibnamefont {Campbell}}, \bibinfo {author} {\bibfnamefont {Y.}~\bibnamefont {Chen}}, \bibinfo {author} {\bibfnamefont {Z.}~\bibnamefont {Chen}}, \bibinfo {author} {\bibfnamefont {B.}~\bibnamefont {Chiaro}}, \bibinfo {author} {\bibfnamefont {A.}~\bibnamefont {Dunsworth}}, \bibinfo {author} {\bibfnamefont {A.~G.}\
  \bibnamefont {Fowler}}, \bibinfo {author} {\bibfnamefont {E.}~\bibnamefont {Jeffrey}}, \bibinfo {author} {\bibfnamefont {E.}~\bibnamefont {Lucero}}, \bibinfo {author} {\bibfnamefont {A.}~\bibnamefont {Megrant}}, \bibinfo {author} {\bibfnamefont {J.~Y.}\ \bibnamefont {Mutus}}, \bibinfo {author} {\bibfnamefont {M.}~\bibnamefont {Neeley}}, \bibinfo {author} {\bibfnamefont {C.}~\bibnamefont {Neill}}, \bibinfo {author} {\bibfnamefont {C.}~\bibnamefont {Quintana}}, \bibinfo {author} {\bibfnamefont {D.}~\bibnamefont {Sank}}, \bibinfo {author} {\bibfnamefont {A.}~\bibnamefont {Vainsencher}}, \bibinfo {author} {\bibfnamefont {J.}~\bibnamefont {Wenner}}, \bibinfo {author} {\bibfnamefont {T.~C.}\ \bibnamefont {White}}, \bibinfo {author} {\bibfnamefont {P.~V.}\ \bibnamefont {Coveney}}, \bibinfo {author} {\bibfnamefont {P.~J.}\ \bibnamefont {Love}}, \bibinfo {author} {\bibfnamefont {H.}~\bibnamefont {Neven}}, \bibinfo {author} {\bibfnamefont {A.}~\bibnamefont {Aspuru-Guzik}},\ and\ \bibinfo {author} {\bibfnamefont
  {J.~M.}\ \bibnamefont {Martinis}},\ }\bibfield  {title} {\bibinfo {title} {Scalable quantum simulation of molecular energies},\ }\href {https://doi.org/10.1103/PhysRevX.6.031007} {\bibfield  {journal} {\bibinfo  {journal} {Phys. Rev. X}\ }\textbf {\bibinfo {volume} {6}},\ \bibinfo {pages} {031007} (\bibinfo {year} {2016})}\BibitemShut {NoStop}%
\bibitem [{\citenamefont {Childs}\ and\ \citenamefont {Su}(2019)}]{childs-prl2019-pf}%
  \BibitemOpen
  \bibfield  {author} {\bibinfo {author} {\bibfnamefont {A.~M.}\ \bibnamefont {Childs}}\ and\ \bibinfo {author} {\bibfnamefont {Y.}~\bibnamefont {Su}},\ }\bibfield  {title} {\bibinfo {title} {Nearly optimal lattice simulation by product formulas},\ }\href {https://doi.org/10.1103/PhysRevLett.123.050503} {\bibfield  {journal} {\bibinfo  {journal} {Phys. Rev. Lett.}\ }\textbf {\bibinfo {volume} {123}},\ \bibinfo {pages} {050503} (\bibinfo {year} {2019})}\BibitemShut {NoStop}%
\bibitem [{\citenamefont {Childs}\ \emph {et~al.}(2021)\citenamefont {Childs}, \citenamefont {Su}, \citenamefont {Tran}, \citenamefont {Wiebe},\ and\ \citenamefont {Zhu}}]{childs2021-trotter}%
  \BibitemOpen
  \bibfield  {author} {\bibinfo {author} {\bibfnamefont {A.~M.}\ \bibnamefont {Childs}}, \bibinfo {author} {\bibfnamefont {Y.}~\bibnamefont {Su}}, \bibinfo {author} {\bibfnamefont {M.~C.}\ \bibnamefont {Tran}}, \bibinfo {author} {\bibfnamefont {N.}~\bibnamefont {Wiebe}},\ and\ \bibinfo {author} {\bibfnamefont {S.}~\bibnamefont {Zhu}},\ }\bibfield  {title} {\bibinfo {title} {Theory of trotter error with commutator scaling},\ }\href {https://doi.org/10.1103/PhysRevX.11.011020} {\bibfield  {journal} {\bibinfo  {journal} {Phys. Rev. X}\ }\textbf {\bibinfo {volume} {11}},\ \bibinfo {pages} {011020} (\bibinfo {year} {2021})}\BibitemShut {NoStop}%
\bibitem [{\citenamefont {Zhao}\ \emph {et~al.}(2022)\citenamefont {Zhao}, \citenamefont {Zhou}, \citenamefont {Shaw}, \citenamefont {Li},\ and\ \citenamefont {Childs}}]{Zhao-prl2022-random}%
  \BibitemOpen
  \bibfield  {author} {\bibinfo {author} {\bibfnamefont {Q.}~\bibnamefont {Zhao}}, \bibinfo {author} {\bibfnamefont {Y.}~\bibnamefont {Zhou}}, \bibinfo {author} {\bibfnamefont {A.~F.}\ \bibnamefont {Shaw}}, \bibinfo {author} {\bibfnamefont {T.}~\bibnamefont {Li}},\ and\ \bibinfo {author} {\bibfnamefont {A.~M.}\ \bibnamefont {Childs}},\ }\bibfield  {title} {\bibinfo {title} {Hamiltonian simulation with random inputs},\ }\href {https://doi.org/10.1103/PhysRevLett.129.270502} {\bibfield  {journal} {\bibinfo  {journal} {Phys. Rev. Lett.}\ }\textbf {\bibinfo {volume} {129}},\ \bibinfo {pages} {270502} (\bibinfo {year} {2022})}\BibitemShut {NoStop}%
\bibitem [{\citenamefont {Zhao}\ \emph {et~al.}(2024)\citenamefont {Zhao}, \citenamefont {Zhou},\ and\ \citenamefont {Childs}}]{zhao2024-entanglement}%
  \BibitemOpen
  \bibfield  {author} {\bibinfo {author} {\bibfnamefont {Q.}~\bibnamefont {Zhao}}, \bibinfo {author} {\bibfnamefont {Y.}~\bibnamefont {Zhou}},\ and\ \bibinfo {author} {\bibfnamefont {A.~M.}\ \bibnamefont {Childs}},\ }\bibfield  {title} {\bibinfo {title} {Entanglement accelerates quantum simulation},\ }\Eprint {https://arxiv.org/abs/2406.02379} {arXiv:2406.02379 [quant-ph]}  (\bibinfo {year} {2024})\BibitemShut {NoStop}%
\bibitem [{\citenamefont {Low}\ and\ \citenamefont {Chuang}(2017)}]{Low2017-ig-low-energy}%
  \BibitemOpen
  \bibfield  {author} {\bibinfo {author} {\bibfnamefont {G.~H.}\ \bibnamefont {Low}}\ and\ \bibinfo {author} {\bibfnamefont {I.~L.}\ \bibnamefont {Chuang}},\ }\bibfield  {title} {\bibinfo {title} {Hamiltonian simulation by uniform spectral amplification},\ }\Eprint {https://arxiv.org/abs/1707.05391} {1707.05391}  (\bibinfo {year} {2017})\BibitemShut {NoStop}%
\bibitem [{\citenamefont {Gu}\ \emph {et~al.}(2021)\citenamefont {Gu}, \citenamefont {Somma},\ and\ \citenamefont {Şahinoğlu}}]{Gu_2021-low-energy}%
  \BibitemOpen
  \bibfield  {author} {\bibinfo {author} {\bibfnamefont {S.}~\bibnamefont {Gu}}, \bibinfo {author} {\bibfnamefont {R.~D.}\ \bibnamefont {Somma}},\ and\ \bibinfo {author} {\bibfnamefont {B.}~\bibnamefont {Şahinoğlu}},\ }\bibfield  {title} {\bibinfo {title} {Fast-forwarding quantum evolution},\ }\href {https://doi.org/10.22331/q-2021-11-15-577} {\bibfield  {journal} {\bibinfo  {journal} {Quantum}\ }\textbf {\bibinfo {volume} {5}},\ \bibinfo {pages} {577} (\bibinfo {year} {2021})}\BibitemShut {NoStop}%
\bibitem [{\citenamefont {Zlokapa}\ and\ \citenamefont {Somma}(2024)}]{Zlokapa_2024-low-energy}%
  \BibitemOpen
  \bibfield  {author} {\bibinfo {author} {\bibfnamefont {A.}~\bibnamefont {Zlokapa}}\ and\ \bibinfo {author} {\bibfnamefont {R.~D.}\ \bibnamefont {Somma}},\ }\bibfield  {title} {\bibinfo {title} {Hamiltonian simulation for low-energy states with optimal time dependence},\ }\href {https://doi.org/10.22331/q-2024-08-27-1449} {\bibfield  {journal} {\bibinfo  {journal} {Quantum}\ }\textbf {\bibinfo {volume} {8}},\ \bibinfo {pages} {1449} (\bibinfo {year} {2024})}\BibitemShut {NoStop}%
\bibitem [{\citenamefont {{\c S}ahino{\u g}lu}\ and\ \citenamefont {Somma}(2021)}]{Sahinoglu2021-ng-low-energy}%
  \BibitemOpen
  \bibfield  {author} {\bibinfo {author} {\bibfnamefont {B.}~\bibnamefont {{\c S}ahino{\u g}lu}}\ and\ \bibinfo {author} {\bibfnamefont {R.~D.}\ \bibnamefont {Somma}},\ }\bibfield  {title} {\bibinfo {title} {Hamiltonian simulation in the low-energy subspace},\ }\href {https://doi.org/https://doi.org/10.1038/s41534-021-00451-w} {\bibfield  {journal} {\bibinfo  {journal} {npj Quantum Information}\ }\textbf {\bibinfo {volume} {7}},\ \bibinfo {pages} {1} (\bibinfo {year} {2021})}\BibitemShut {NoStop}%
\bibitem [{\citenamefont {Gong}\ \emph {et~al.}(2024)\citenamefont {Gong}, \citenamefont {Zhou},\ and\ \citenamefont {Li}}]{Gong2024-wr-low-energy}%
  \BibitemOpen
  \bibfield  {author} {\bibinfo {author} {\bibfnamefont {W.}~\bibnamefont {Gong}}, \bibinfo {author} {\bibfnamefont {S.}~\bibnamefont {Zhou}},\ and\ \bibinfo {author} {\bibfnamefont {T.}~\bibnamefont {Li}},\ }\bibfield  {title} {\bibinfo {title} {Complexity of digital quantum simulation in the low-energy subspace: Applications and a lower bound},\ }\href {https://doi.org/https://doi.org/10.22331/q-2024-07-15-1409} {\bibfield  {journal} {\bibinfo  {journal} {Quantum}\ }\textbf {\bibinfo {volume} {8}},\ \bibinfo {pages} {1409} (\bibinfo {year} {2024})}\BibitemShut {NoStop}%
\bibitem [{\citenamefont {Hejazi}\ \emph {et~al.}(2024{\natexlab{a}})\citenamefont {Hejazi}, \citenamefont {Zini},\ and\ \citenamefont {Arrazola}}]{Hejazi2024-xs-low-energy}%
  \BibitemOpen
  \bibfield  {author} {\bibinfo {author} {\bibfnamefont {K.}~\bibnamefont {Hejazi}}, \bibinfo {author} {\bibfnamefont {M.~S.}\ \bibnamefont {Zini}},\ and\ \bibinfo {author} {\bibfnamefont {J.~M.}\ \bibnamefont {Arrazola}},\ }\bibfield  {title} {\bibinfo {title} {Better bounds for low-energy product formulas},\ }\Eprint {https://arxiv.org/abs/2402.10362} {arXiv:2402.10362 [quant-ph]}  (\bibinfo {year} {2024}{\natexlab{a}})\BibitemShut {NoStop}%
\bibitem [{\citenamefont {Hejazi}\ \emph {et~al.}(2024{\natexlab{b}})\citenamefont {Hejazi}, \citenamefont {Soni}, \citenamefont {Zini},\ and\ \citenamefont {Arrazola}}]{hejazi-2024arXiv-low-energy-qpe}%
  \BibitemOpen
  \bibfield  {author} {\bibinfo {author} {\bibfnamefont {K.}~\bibnamefont {Hejazi}}, \bibinfo {author} {\bibfnamefont {J.}~\bibnamefont {Soni}}, \bibinfo {author} {\bibfnamefont {M.~S.}\ \bibnamefont {Zini}},\ and\ \bibinfo {author} {\bibfnamefont {J.~M.}\ \bibnamefont {Arrazola}},\ }\bibfield  {title} {\bibinfo {title} {Better product formulas for quantum phase estimation},\ }\Eprint {https://arxiv.org/abs/2412.16811} {arXiv:2412.16811 [quant-ph]}  (\bibinfo {year} {2024}{\natexlab{b}})\BibitemShut {NoStop}%
\bibitem [{Sup()}]{Supple}%
  \BibitemOpen
  \href@noop {} {}\bibinfo {note} {See Supplementary Materials for the detailed discussion. Section \ref{SecS:num_terms} gives the number of terms in Trotterization. Sections \ref{SecS:nested_com} and \ref{SecS:trotter_error} are devoted to the detailed derivation of Theorem \ref{Thm:Error_low_energy}. Section \ref{SecS:trotter_number} gives the exact scaling of the Trotter number, Eq. (\ref{Eq:trotter_number}). Section \ref{SecS:Numerical} shows the numerical demonstration of Theorem \ref{Thm:Error_low_energy}. Section \ref{SecS:weakly_correlated} gives an extension of our result to weakly-correlated states with low-energy expectation values.}\BibitemShut {Stop}%
\bibitem [{\citenamefont {Arad}\ \emph {et~al.}(2016)\citenamefont {Arad}, \citenamefont {Kuwahara},\ and\ \citenamefont {Landau}}]{Arad2016-ak-excitation}%
  \BibitemOpen
  \bibfield  {author} {\bibinfo {author} {\bibfnamefont {I.}~\bibnamefont {Arad}}, \bibinfo {author} {\bibfnamefont {T.}~\bibnamefont {Kuwahara}},\ and\ \bibinfo {author} {\bibfnamefont {Z.}~\bibnamefont {Landau}},\ }\bibfield  {title} {\bibinfo {title} {Connecting global and local energy distributions in quantum spin models on a lattice},\ }\href {https://doi.org/10.1088/1742-5468/2016/03/033301} {\bibfield  {journal} {\bibinfo  {journal} {J. Stat. Mech.}\ }\textbf {\bibinfo {volume} {2016}},\ \bibinfo {pages} {033301} (\bibinfo {year} {2016})}\BibitemShut {NoStop}%
\bibitem [{\citenamefont {Anshu}(2016)}]{Anshu2016-gc-concentration}%
  \BibitemOpen
  \bibfield  {author} {\bibinfo {author} {\bibfnamefont {A.}~\bibnamefont {Anshu}},\ }\bibfield  {title} {\bibinfo {title} {Concentration bounds for quantum states with finite correlation length on quantum spin lattice systems},\ }\href {https://doi.org/10.1088/1367-2630/18/8/083011} {\bibfield  {journal} {\bibinfo  {journal} {New J. Phys.}\ }\textbf {\bibinfo {volume} {18}},\ \bibinfo {pages} {083011} (\bibinfo {year} {2016})}\BibitemShut {NoStop}%
\bibitem [{\citenamefont {Kuwahara}(2016)}]{Kuwahara2016-gh-Chernoff}%
  \BibitemOpen
  \bibfield  {author} {\bibinfo {author} {\bibfnamefont {T.}~\bibnamefont {Kuwahara}},\ }\bibfield  {title} {\bibinfo {title} {Connecting the probability distributions of different operators and generalization of the {Chernoff–Hoeffding} inequality},\ }\href {https://doi.org/10.1088/1742-5468/2016/11/113103} {\bibfield  {journal} {\bibinfo  {journal} {J. Stat. Mech.}\ }\textbf {\bibinfo {volume} {2016}},\ \bibinfo {pages} {113103} (\bibinfo {year} {2016})}\BibitemShut {NoStop}%
\bibitem [{\citenamefont {Campbell}(2019)}]{Campbell-prl2019-qdrift}%
  \BibitemOpen
  \bibfield  {author} {\bibinfo {author} {\bibfnamefont {E.}~\bibnamefont {Campbell}},\ }\bibfield  {title} {\bibinfo {title} {Random compiler for fast hamiltonian simulation},\ }\href {https://doi.org/10.1103/PhysRevLett.123.070503} {\bibfield  {journal} {\bibinfo  {journal} {Phys. Rev. Lett.}\ }\textbf {\bibinfo {volume} {123}},\ \bibinfo {pages} {070503} (\bibinfo {year} {2019})}\BibitemShut {NoStop}%
\bibitem [{\citenamefont {Low}\ \emph {et~al.}(2019)\citenamefont {Low}, \citenamefont {Kliuchnikov},\ and\ \citenamefont {Wiebe}}]{low2019-mpf}%
  \BibitemOpen
  \bibfield  {author} {\bibinfo {author} {\bibfnamefont {G.~H.}\ \bibnamefont {Low}}, \bibinfo {author} {\bibfnamefont {V.}~\bibnamefont {Kliuchnikov}},\ and\ \bibinfo {author} {\bibfnamefont {N.}~\bibnamefont {Wiebe}},\ }\bibfield  {title} {\bibinfo {title} {{Well-conditioned multiproduct Hamiltonian simulation}},\ }\Eprint {https://arxiv.org/abs/1907.11679} {arXiv:1907.11679 [quant-ph]}  (\bibinfo {year} {2019})\BibitemShut {NoStop}%
\bibitem [{\citenamefont {Zhuk}\ \emph {et~al.}(2024)\citenamefont {Zhuk}, \citenamefont {Robertson},\ and\ \citenamefont {Bravyi}}]{zhuk2024-mpf}%
  \BibitemOpen
  \bibfield  {author} {\bibinfo {author} {\bibfnamefont {S.}~\bibnamefont {Zhuk}}, \bibinfo {author} {\bibfnamefont {N.}~\bibnamefont {Robertson}},\ and\ \bibinfo {author} {\bibfnamefont {S.}~\bibnamefont {Bravyi}},\ }\bibfield  {title} {\bibinfo {title} {{Trotter error bounds and dynamic multi-product formulas for Hamiltonian simulation}},\ }\Eprint {https://arxiv.org/abs/2306.12569} {arXiv:2306.12569 [quant-ph]}  (\bibinfo {year} {2024})\BibitemShut {NoStop}%
\bibitem [{\citenamefont {Aftab}\ \emph {et~al.}(2024)\citenamefont {Aftab}, \citenamefont {An},\ and\ \citenamefont {Trivisa}}]{aftab2024-mpf}%
  \BibitemOpen
  \bibfield  {author} {\bibinfo {author} {\bibfnamefont {J.}~\bibnamefont {Aftab}}, \bibinfo {author} {\bibfnamefont {D.}~\bibnamefont {An}},\ and\ \bibinfo {author} {\bibfnamefont {K.}~\bibnamefont {Trivisa}},\ }\bibfield  {title} {\bibinfo {title} {{Multi-product Hamiltonian simulation with explicit commutator scaling}},\ }\Eprint {https://arxiv.org/abs/2403.08922} {arXiv:2403.08922 [quant-ph]}  (\bibinfo {year} {2024})\BibitemShut {NoStop}%
\bibitem [{\citenamefont {Watson}\ and\ \citenamefont {Watkins}(2024)}]{Watson2024-mpf}%
  \BibitemOpen
  \bibfield  {author} {\bibinfo {author} {\bibfnamefont {J.~D.}\ \bibnamefont {Watson}}\ and\ \bibinfo {author} {\bibfnamefont {J.}~\bibnamefont {Watkins}},\ }\bibfield  {title} {\bibinfo {title} {{Exponentially Reduced Circuit Depths Using Trotter Error Mitigation}},\ }\Eprint {https://arxiv.org/abs/2408.14385} {arXiv:2408.14385 [quant-ph]}  (\bibinfo {year} {2024})\BibitemShut {NoStop}%
\bibitem [{\citenamefont {Suzuki}(1993)}]{Suzuki1993-general}%
  \BibitemOpen
  \bibfield  {author} {\bibinfo {author} {\bibfnamefont {M.}~\bibnamefont {Suzuki}},\ }\bibfield  {title} {\bibinfo {title} {General decomposition theory of ordered exponentials},\ }\href {https://doi.org/10.2183/PJAB.69.161} {\bibfield  {journal} {\bibinfo  {journal} {Proceedings of the Japan Academy, Series B}\ }\textbf {\bibinfo {volume} {69}},\ \bibinfo {pages} {161} (\bibinfo {year} {1993})}\BibitemShut {NoStop}%
\bibitem [{\citenamefont {Wiebe}\ \emph {et~al.}(2010)\citenamefont {Wiebe}, \citenamefont {Berry}, \citenamefont {H{\o}yer},\ and\ \citenamefont {Sanders}}]{Wiebe2010-mu}%
  \BibitemOpen
  \bibfield  {author} {\bibinfo {author} {\bibfnamefont {N.}~\bibnamefont {Wiebe}}, \bibinfo {author} {\bibfnamefont {D.}~\bibnamefont {Berry}}, \bibinfo {author} {\bibfnamefont {P.}~\bibnamefont {H{\o}yer}},\ and\ \bibinfo {author} {\bibfnamefont {B.~C.}\ \bibnamefont {Sanders}},\ }\bibfield  {title} {\bibinfo {title} {Higher order decompositions of ordered operator exponentials},\ }\href {https://doi.org/10.1088/1751-8113/43/6/065203} {\bibfield  {journal} {\bibinfo  {journal} {J. Phys. A: Math. Theor.}\ }\textbf {\bibinfo {volume} {43}},\ \bibinfo {pages} {065203} (\bibinfo {year} {2010})}\BibitemShut {NoStop}%
\bibitem [{\citenamefont {Mizuta}\ \emph {et~al.}(2024)\citenamefont {Mizuta}, \citenamefont {Ikeda},\ and\ \citenamefont {Fujii}}]{mizuta2024arxiv-time-dep-PF}%
  \BibitemOpen
  \bibfield  {author} {\bibinfo {author} {\bibfnamefont {K.}~\bibnamefont {Mizuta}}, \bibinfo {author} {\bibfnamefont {T.~N.}\ \bibnamefont {Ikeda}},\ and\ \bibinfo {author} {\bibfnamefont {K.}~\bibnamefont {Fujii}},\ }\bibfield  {title} {\bibinfo {title} {Explicit error bounds with commutator scaling for time-dependent product and multi-product formulas},\ }\Eprint {https://arxiv.org/abs/2410.14243} {arXiv:2410.14243 [quant-ph]}  (\bibinfo {year} {2024})\BibitemShut {NoStop}%
\bibitem [{\citenamefont {Cao}\ \emph {et~al.}(2024)\citenamefont {Cao}, \citenamefont {Jin},\ and\ \citenamefont {Liu}}]{cao2024arxiv-time-dep-PF}%
  \BibitemOpen
  \bibfield  {author} {\bibinfo {author} {\bibfnamefont {Y.}~\bibnamefont {Cao}}, \bibinfo {author} {\bibfnamefont {S.}~\bibnamefont {Jin}},\ and\ \bibinfo {author} {\bibfnamefont {N.}~\bibnamefont {Liu}},\ }\bibfield  {title} {\bibinfo {title} {A unifying framework for quantum simulation algorithms for time-dependent hamiltonian dynamics},\ }\Eprint {https://arxiv.org/abs/2411.03180} {arXiv:2411.03180 [quant-ph]}  (\bibinfo {year} {2024})\BibitemShut {NoStop}%
\bibitem [{\citenamefont {Aspuru-Guzik}\ \emph {et~al.}(2005)\citenamefont {Aspuru-Guzik}, \citenamefont {Dutoi}, \citenamefont {Love},\ and\ \citenamefont {Head-Gordon}}]{Aspuru-Guzik2005-adiabatic}%
  \BibitemOpen
  \bibfield  {author} {\bibinfo {author} {\bibfnamefont {A.}~\bibnamefont {Aspuru-Guzik}}, \bibinfo {author} {\bibfnamefont {A.~D.}\ \bibnamefont {Dutoi}}, \bibinfo {author} {\bibfnamefont {P.~J.}\ \bibnamefont {Love}},\ and\ \bibinfo {author} {\bibfnamefont {M.}~\bibnamefont {Head-Gordon}},\ }\bibfield  {title} {\bibinfo {title} {Simulated quantum computation of molecular energies},\ }\href {https://doi.org/10.1126/science.1113479} {\bibfield  {journal} {\bibinfo  {journal} {Science}\ }\textbf {\bibinfo {volume} {309}},\ \bibinfo {pages} {1704} (\bibinfo {year} {2005})}\BibitemShut {NoStop}%
\bibitem [{\citenamefont {Du}\ \emph {et~al.}(2010)\citenamefont {Du}, \citenamefont {Xu}, \citenamefont {Peng}, \citenamefont {Wang}, \citenamefont {Wu},\ and\ \citenamefont {Lu}}]{Du-prl2010-adiabatic}%
  \BibitemOpen
  \bibfield  {author} {\bibinfo {author} {\bibfnamefont {J.}~\bibnamefont {Du}}, \bibinfo {author} {\bibfnamefont {N.}~\bibnamefont {Xu}}, \bibinfo {author} {\bibfnamefont {X.}~\bibnamefont {Peng}}, \bibinfo {author} {\bibfnamefont {P.}~\bibnamefont {Wang}}, \bibinfo {author} {\bibfnamefont {S.}~\bibnamefont {Wu}},\ and\ \bibinfo {author} {\bibfnamefont {D.}~\bibnamefont {Lu}},\ }\bibfield  {title} {\bibinfo {title} {Nmr implementation of a molecular hydrogen quantum simulation with adiabatic state preparation},\ }\href {https://doi.org/10.1103/PhysRevLett.104.030502} {\bibfield  {journal} {\bibinfo  {journal} {Phys. Rev. Lett.}\ }\textbf {\bibinfo {volume} {104}},\ \bibinfo {pages} {030502} (\bibinfo {year} {2010})}\BibitemShut {NoStop}%
\bibitem [{\citenamefont {Babbush}\ \emph {et~al.}(2018)\citenamefont {Babbush}, \citenamefont {Wiebe}, \citenamefont {McClean}, \citenamefont {McClain}, \citenamefont {Neven},\ and\ \citenamefont {Chan}}]{Babbush-2018-FFFT}%
  \BibitemOpen
  \bibfield  {author} {\bibinfo {author} {\bibfnamefont {R.}~\bibnamefont {Babbush}}, \bibinfo {author} {\bibfnamefont {N.}~\bibnamefont {Wiebe}}, \bibinfo {author} {\bibfnamefont {J.}~\bibnamefont {McClean}}, \bibinfo {author} {\bibfnamefont {J.}~\bibnamefont {McClain}}, \bibinfo {author} {\bibfnamefont {H.}~\bibnamefont {Neven}},\ and\ \bibinfo {author} {\bibfnamefont {G.~K.-L.}\ \bibnamefont {Chan}},\ }\bibfield  {title} {\bibinfo {title} {Low-depth quantum simulation of materials},\ }\href {https://doi.org/10.1103/PhysRevX.8.011044} {\bibfield  {journal} {\bibinfo  {journal} {Phys. Rev. X}\ }\textbf {\bibinfo {volume} {8}},\ \bibinfo {pages} {011044} (\bibinfo {year} {2018})}\BibitemShut {NoStop}%
\bibitem [{\citenamefont {Kuwahara}\ \emph {et~al.}(2016)\citenamefont {Kuwahara}, \citenamefont {Mori},\ and\ \citenamefont {Saito}}]{Kuwahara2016-yn}%
  \BibitemOpen
  \bibfield  {author} {\bibinfo {author} {\bibfnamefont {T.}~\bibnamefont {Kuwahara}}, \bibinfo {author} {\bibfnamefont {T.}~\bibnamefont {Mori}},\ and\ \bibinfo {author} {\bibfnamefont {K.}~\bibnamefont {Saito}},\ }\bibfield  {title} {\bibinfo {title} {{Floquet–Magnus} theory and generic transient dynamics in periodically driven many-body quantum systems},\ }\href {https://doi.org/https://doi.org/10.1016/j.aop.2016.01.012} {\bibfield  {journal} {\bibinfo  {journal} {Ann. Phys.}\ }\textbf {\bibinfo {volume} {367}},\ \bibinfo {pages} {96} (\bibinfo {year} {2016})}\BibitemShut {NoStop}%
\bibitem [{\citenamefont {Affleck}\ \emph {et~al.}(1988)\citenamefont {Affleck}, \citenamefont {Kennedy}, \citenamefont {Lieb},\ and\ \citenamefont {Tasaki}}]{Affleck1988-ia}%
  \BibitemOpen
  \bibfield  {author} {\bibinfo {author} {\bibfnamefont {I.}~\bibnamefont {Affleck}}, \bibinfo {author} {\bibfnamefont {T.}~\bibnamefont {Kennedy}}, \bibinfo {author} {\bibfnamefont {E.~H.}\ \bibnamefont {Lieb}},\ and\ \bibinfo {author} {\bibfnamefont {H.}~\bibnamefont {Tasaki}},\ }\bibfield  {title} {\bibinfo {title} {Valence bond ground states in isotropic quantum antiferromagnets},\ }\href {https://doi.org/10.1007/bf01218021} {\bibfield  {journal} {\bibinfo  {journal} {Commun. Math. Phys.}\ }\textbf {\bibinfo {volume} {115}},\ \bibinfo {pages} {477} (\bibinfo {year} {1988})}\BibitemShut {NoStop}%
\bibitem [{\citenamefont {Majumdar}\ and\ \citenamefont {Ghosh}(1969{\natexlab{a}})}]{Majumdar1969-cn-1}%
  \BibitemOpen
  \bibfield  {author} {\bibinfo {author} {\bibfnamefont {C.~K.}\ \bibnamefont {Majumdar}}\ and\ \bibinfo {author} {\bibfnamefont {D.~K.}\ \bibnamefont {Ghosh}},\ }\bibfield  {title} {\bibinfo {title} {On next‐nearest‐neighbor interaction in linear chain. {I}},\ }\href {https://doi.org/https://doi.org/10.1063/1.1664978} {\bibfield  {journal} {\bibinfo  {journal} {Journal of Mathematical Physics}\ }\textbf {\bibinfo {volume} {10}},\ \bibinfo {pages} {1388} (\bibinfo {year} {1969}{\natexlab{a}})}\BibitemShut {NoStop}%
\bibitem [{\citenamefont {Majumdar}\ and\ \citenamefont {Ghosh}(1969{\natexlab{b}})}]{Majumdar1969-rw-2}%
  \BibitemOpen
  \bibfield  {author} {\bibinfo {author} {\bibfnamefont {C.~K.}\ \bibnamefont {Majumdar}}\ and\ \bibinfo {author} {\bibfnamefont {D.~K.}\ \bibnamefont {Ghosh}},\ }\bibfield  {title} {\bibinfo {title} {On next‐nearest‐neighbor interaction in linear chain. {II}},\ }\href {https://doi.org/https://doi.org/10.1063/1.1664979} {\bibfield  {journal} {\bibinfo  {journal} {Journal of Mathematical Physics}\ }\textbf {\bibinfo {volume} {10}},\ \bibinfo {pages} {1399} (\bibinfo {year} {1969}{\natexlab{b}})}\BibitemShut {NoStop}%
\end{thebibliography}%

\clearpage
\onecolumngrid

\begin{center}
\textbf{\large Supplemental Materials for \\``Trotterization is substantially efficient for low-energy states"}
\end{center}
\begin{center}
    Kaoru Mizuta$^{1,2,3}$, Tomotaka Kuwahara$^{4,5,6}$ \\
    \vspace{5pt}
    {\small \textit{$^1$ Department of Applied Physics, Graduate School of Engineering, \\ The University of Tokyo, Hongo 7-3-1, Bunkyo, Tokyo 113-8656, Japan}} \\
   {\small \textit{$^2$ Photon Science Center, Graduate School of Engineering, \\ The University of Tokyo, Hongo 7-3-1, Bunkyo, Tokyo 113-8656, Japan}} \\
   {\small \textit{$^3$ RIKEN Center for Quantum Computing (RQC), Hirosawa 2-1, Wako, Saitama 351-0198, Japan}} \\
   {\small \textit{$^4$ Analytical Quantum Complexity RIKEN Hakubi Research Team, RIKEN Center for Quantum Computing (RQC), Wako, Saitama 351-0198, Japan}} \\
  {\small \textit{$^5$ RIKEN Cluster for Pioneering Research (CPR), Wako, Saitama 351-0198, Japan}} \\
  {\small \textit{$^6$ PRESTO, Japan Science and Technology (JST), Kawaguchi, Saitama 332-0012, Japan }}
   (Dated: \today)
\end{center}

\renewcommand{\thesection}{S\arabic{section}}
\renewcommand{\theequation}{S\arabic{equation}}
\setcounter{equation}{0}
\renewcommand{\thetheorem}{S\arabic{theorem}}
\setcounter{theorem}{0}

\renewcommand{\thefigure}{S\arabic{figure}}

\section{Number of terms in Trotterization}\label{SecS:num_terms}

Trotterization requires decomposition of a Hamiltonian,
\begin{equation}
    H = \sum_{\gamma=1}^\Gamma H_\gamma, \quad H_\gamma = \sum_{X \subset \Lambda} h_X^\gamma,
\end{equation}
where each pair of $\{h_X^\gamma\}_X$ commutes as $[h_X^\gamma,h_{X'}^\gamma] = 0$.
Since the number of partition $\Gamma$ is a significant factor in our result, we briefly mention how it scales for various classes of quantum many-body Hamiltonians.
We focus on whether $\Gamma$ is $\order{1}$ or not.

\textbf{Local Hamiltonians with finite-ranged interactions.---}
Local Hamiltonians with finite-ranged interactions are typical examples satifying $\Gamma \in \order{1}$.
For instance, suppose that a Hamiltonian has nearest neighbor interactions on a one-dimensional lattice as
\begin{equation}
    H = \sum_{i=1}^{N-1} h_{i,i+1}, \quad h_{i,i+1} \geq 0.
\end{equation}
It can be decomposed into two terms $H_1 = \sum_{i;\text{odd}} h_{i,i+1}$ and $H_2 = \sum_{i;\text{even}} h_{i,i+1}$, where $\Gamma$ is equal to $2$.
For any locality $k \in \order{1}$ and any spatial dimension $d \in \order{1}$, the number $\Gamma$ can be $\order{1}$ as well.

\textbf{Local Hamiltonians with long-ranged interactions composed of commuting operators.---}
Some of local Hamiltonians with long-ranged interactions can satisfy $\Gamma \in \order{1}$.
A useful class is the case where the interactions can be classified into some groups having mutually commuting operators.
To be concrete, let us consider a Hamiltonian having long-ranged Heisenberg interactions,
\begin{equation}
    H = \sum_{i,j \in \Lambda} \left(J_{ij}^x \frac{X_i X_j+1}{2} + J_{ij}^y \frac{Y_i Y_j+1}{2} + J_{ij}^z \frac{Z_i Z_j+1}{2} \right).
\end{equation}
Then, we can decompose it with $\Gamma=3$ by the following three terms, $H_1 = \sum_{i,j \in \Lambda} J_{ij}^x \frac{X_i X_j+1}{2}$, $H_2 = \sum_{i,j \in \Lambda} J_{ij}^y \frac{Y_i Y_j+1}{2}$, and $H_3 = \sum_{i,j \in \Lambda} J_{ij}^z \frac{Z_i Z_j+1}{2}$.
Another example in this class is a fermionic Hamiltonian,
\begin{equation}
    H = \sum_k \varepsilon(k) \hat{c}_k^\dagger \hat{c}_k + \sum_{i,j \in \Lambda} U_{ij} \hat{n}_i \hat{n}_j,
\end{equation}
where $\hat{c}_k$ and $\hat{c}_k^\dagger$ are fermionic annihilation and creation operators in the wave number space, and $\hat{n}_j$ is a number operator in the real space.
The decomposition by $H_1 = \sum_k \varepsilon(k) \hat{c}_k^\dagger \hat{c}_k$ and $H_2 = \sum_{i,j \in \Lambda} U_{ij} \hat{n}_i \hat{n}_j$ results in $\Gamma =2$, where the time-evolution under $H_1$ is implemented by the fast fermionic Fourier transform (FFFT) \cite{Babbush-2018-FFFT}.
Various models with any locality and dimension belongs to this class. 

\textbf{Other local Hamiltonians.---}
In general, any local Hamiltonian with $k \in \order{1}$ can be decomposed into $\Gamma \in \order{N^k}$ terms with choosing $h_X$ as each term $H_\gamma$.
We use our results for $\Gamma \notin \order{1}$ in such cases.

\section{Nested commutators in the low-energy subspace}\label{SecS:nested_com}

Here, we prove the upper bound on the nested commutators in the low-energy subspace as follows.

\begin{theorem}\label{ThmS:nested_com_low_energy}
\textbf{(Nested commutators in the low-energy subspace)}

Let $H_{q'} = \sum_{X \subset \Lambda} h_X^{q'}$ be $k$-local and $g$-extensive Hamiltonians with positive-semidefinite terms $h_X^{q'} \geq 0$.
When there exists a projection $\Pi$ such that $\Pi H_{q'} \Pi \leq \Delta$ for every $q' = 0,1,\cdots,q$, the nested commutator in the subspace designated by the projection $\Pi$ is bounded by
\begin{equation}\label{EqS:nested_com_low_proj}
    \norm{\Pi [H_{q},[H_{q-1},\cdots,[H_1,H_0]]] \Pi} \leq q! (2kg)^q \Delta.
\end{equation}
\end{theorem}

The nested commutator among $k$-local and $g$-extensive Hamiltonians $H_0,\cdots,H_q$ is known to generally have the upper bound \cite{Kuwahara2016-yn,childs2021-trotter}, 
\begin{equation}\label{EqS:nested_com_norm_general}
    \norm{[H_q,[H_{q-1},\cdots,[H_1,H_0]]]} \leq q! (2kg)^q Ng.
\end{equation}
Theorem \ref{ThmS:nested_com_low_energy} states that the whole energy scale $Ng$ is replaced by the maximum energy of the subspace, $\Delta$.
We note that, in spite of large difference in the derivation for them, the upper bound for low-energy states completely reproduces the one for arbitrary states including its coefficient and scaling in the Trotter order $p$, the locality $k$, and the extensiveness $g$ at $\Delta=Ng$.

\textbf{Proof of Theorem \ref{ThmS:nested_com_low_energy}.---}
Since the nested commutator is hermitian or anti-hermitian, its operator norm is equal to
\begin{eqnarray}
    \norm{\Pi [H_{q},[H_{q-1},\cdots,[H_1,H_0]]] \Pi} &=& \max_{\ket{\psi}; \braket{\psi|\psi}=1} \left| \bra{\psi} \Pi [H_{q},[H_{q-1},\cdots,[H_1,H_0]]] \Pi \ket{\psi} \right| \nonumber \\
    &=& \max_{\substack{\ket{\psi}; \braket{\psi|\psi}=1 \\ \Pi\ket{\psi}=\ket{\psi}}} \left| \bra{\psi} [H_{q},[H_{q-1},\cdots,[H_1,H_0]]] \ket{\psi} \right|. \label{EqS:Com_expectation}
\end{eqnarray}
In the calculation below, a state $\ket{\psi}$ is supposed to satisfy $\braket{\psi|\psi}=1$ and $\Pi\ket{\psi}=\ket{\psi}$, which lead to
\begin{equation}\label{EqS:H_q_leq_delta}
    \braket{\psi|H_{q'}|\psi} \leq \Delta, \quad q' = 0,1,\cdots,q.
\end{equation}
Let $Y_{q'}$ denote the union of the domains $\{X_{q'} \}$ given by
\begin{equation}
    Y_{q'} = X_0 \cup X_1 \cup \cdots \cup X_{q'-1},
\end{equation}
and then, we define a set of domains in the lattice $\Lambda$ by
\begin{equation}
   \mcl{G}_q = \{ (X_0,\cdots,X_q) \in \Lambda^{q+1} \, | \, ^\forall q' \geq 0, \, |X_{q'}| \leq k \,\, \text{and} \,\, ^\forall q' \geq 1, \, X_{q'} \cap Y_{q'} \neq \phi \}.
\end{equation}
Then, the nested commutator in Eq. (\ref{EqS:Com_expectation}) is evaluated by
\begin{eqnarray}
    \left| \bra{\psi} \Pi [H_{q},[H_{q-1},\cdots,[H_1,H_0]]] \Pi \ket{\psi} \right| &=& \sum_{\substack{X_0,\cdots,X_q \\ (X_0,\cdots,X_q) \in \mcl{G}_q}} \left| \braket{\psi|[h_{X_q}^q,[h_{X_{q-1}}^{q-1},\cdots,[h_{X_1}^1,h_{X_0}^0]]]|\psi}\right| \nonumber \\
    &\leq& 2^q \max_\sigma \sum_{\substack{X_0,\cdots,X_q \\ (X_0,\cdots,X_q) \in \mcl{G}_q}} \left| \braket{\psi|h_{X_q}^q h_{X_{\sigma(q-1)}}^{\sigma(q-1)} \cdots h_{X_{\sigma(0)}}^{\sigma(0)}|\psi}\right|, \label{EqS:Bound_com_exp_prod}
\end{eqnarray}
where $\sigma$ denotes permutation of the indices $\{0,1,\cdots,q-1\}$.
We also omit the superscript $q'$ from $h_{X_q'}^{q'}$ due to its clarity below.
Denoting the index for the operator $h_{X_{\sigma(0)}}$ in the right edge by $r=\sigma(0) \in \{0,\cdots,q-1\}$, the summation in Eq. (\ref{EqS:Bound_com_exp_prod}) is bounded by
\begin{eqnarray}
    && \sum_{\substack{X_0,\cdots,X_q \\ (X_0,\cdots,X_q) \in \mcl{G}_q}} \left| \braket{\psi|h_{X_q}^q h_{X_{\sigma(q-1)}}^{\sigma(q-1)} \cdots h_{X_{\sigma(0)}}^{\sigma(0)}|\psi}\right| \nonumber \\
    && \qquad \leq \sum_{\substack{X_0,\cdots,X_q \\ (X_0,\cdots,X_q) \in \mcl{G}_q}} \norm{\sqrt{h_{X_q}} \ket{\psi}} \norm{\sqrt{h_{X_q}}} \left( \prod_{q'\neq r,q} \norm{\sqrt{h_{X_{q'}}}}\right)^2 \norm{\sqrt{h_{X_r}}} \norm{\sqrt{h_{X_r}} \ket{\psi}} \nonumber \\
    && \qquad \leq \sqrt{\sum_{\substack{X_0,\cdots,X_q \\ (X_0,\cdots,X_q) \in \mcl{G}_q}} \norm{\sqrt{h_{X_q}} \ket{\psi}}^2 \norm{\sqrt{h_{X_r}}}^2 \prod_{q'\neq r,q} \norm{\sqrt{h_{X_{q'}}}}^2 } \sqrt{\sum_{\substack{X_0,\cdots,X_q \\ (X_0,\cdots,X_q) \in \mcl{G}_q}} \norm{\sqrt{h_{X_r}} \ket{\psi}}^2 \norm{\sqrt{h_{X_q}}}^2 \prod_{q'\neq r,q} \norm{\sqrt{h_{X_{q'}}}}^2} \nonumber \\
    && \qquad = \sqrt{\sum_{\substack{X_0,\cdots,X_q \\ (X_0,\cdots,X_q) \in \mcl{G}_q}} \braket{\psi|h_{X_q}|\psi} \prod_{q'\neq q} \norm{h_{X_{q'}}}} \sqrt{\sum_{\substack{X_0,\cdots,X_q \\ (X_0,\cdots,X_q) \in \mcl{G}_q}} \braket{\psi|h_{X_r}|\psi} \prod_{q'\neq r} \norm{h_{X_{q'}}}}.  \label{EqS:Bound_com_exp_sqrt}
\end{eqnarray}
In the first inequality, we use the relation $\norm{A\ket{\psi}} \leq \norm{A} \cdot \norm{\ket{\psi}}$ for an operator $A$ and a state $\ket{\psi}$, and also use the positive-semidefiniteness of $h_X$ for the existence of $\sqrt{h_X}$.
The second inequality comes from the Cauchy-Schwartz inequality.
In the last equality, we use the relation $\norm{\sqrt{A}} = \sqrt{\norm{A}}$ for a positive-semidefinite operator $A$, which is easily checked by the spectral decomposition.

We concentrate on evaluate the upper bound on
\begin{equation}\label{EqS:Bound_com_inner0}
    \sum_{\substack{X_0,\cdots,X_q \\ (X_0,\cdots,X_q) \in \mcl{G}_q}} \braket{\psi|h_{X_r}|\psi} \prod_{q'\neq r} \norm{h_{X_{q'}}}
\end{equation}
for $r \in \{0,1,\cdots,q\}$, which appears in Eq. (\ref{EqS:Bound_com_exp_sqrt}).
We begin with taking the summation over $X_{r+1}, \cdots, X_q$, which leads to
\begin{equation}
    \sum_{\substack{X_0,\cdots,X_q \\ (X_0,\cdots,X_q) \in \mcl{G}_q}} \braket{\psi|h_{X_r}|\psi} \prod_{q'\neq r} \norm{h_{X_{q'}}} = \sum_{\substack{X_0,\cdots,X_r \\ (X_0,\cdots,X_r) \in \mcl{G}_r}} \braket{\psi|h_{X_r}|\psi} \prod_{q'=0}^{r-1} \norm{h_{X_{q'}}} \prod_{q'=r+1}^q\left( \sum_{X_{q'}: X_{q'} \cap Y_{q'} \neq \phi} \norm{h_{X_{q'}}}\right). \label{EqS:Bound_com_inner}
\end{equation}
Each union $Y_{q'}=X_0\cup\cdots\cup X_{q'-1}$ contains at most $q'k$ sites.
In combination with the extensiveness $g$ defined by Eq. (\ref{Eq:def_extensiveness}), we obtain
\begin{eqnarray}
    \sum_{X_{q'}: X_{q'} \cap Y_{q'} \neq \phi} \norm{h_{X_{q'}}} &\leq& \sum_{j \in Y_{q'}} \sum_{X_{q'}: X_{q'} \ni j} \norm{h_{X_{q'}}} \nonumber \\
    &\leq& q'kg. \label{EqS:Yq_extensive}
\end{eqnarray}
Using this inequality, Eq. (\ref{EqS:Bound_com_inner}) is further bounded by
\begin{eqnarray}
    \sum_{\substack{X_0,\cdots,X_q \\ (X_0,\cdots,X_q) \in \mcl{G}_q}} \braket{\psi|h_{X_r}|\psi} \prod_{q'\neq r} \norm{h_{X_{q'}}} &\leq& \frac{q!}{r!} (kg)^{q-r} \sum_{\substack{X_0,\cdots,X_r \\ (X_0,\cdots,X_r) \in \mcl{G}_r}} \braket{\psi|h_{X_r}|\psi} \prod_{q'=0}^{r-1} \norm{h_{X_{q'}}} \nonumber \\
    &=& \frac{q!}{r!} (kg)^{q-r} \sum_{X_r: |X_r| \leq k} \braket{\psi|h_{X_r}|\psi} \sum_{\substack{X_0,\cdots,X_{r-1} \\ (X_0,\cdots,X_{r-1}) \in \mcl{G}_{r-1}, \\ Y_r \cap X_r \neq \phi}} \prod_{q'=0}^{r-1} \norm{h_{X_{q'}}}, \label{EqS:Sr_before}
\end{eqnarray}
where the second equality is the result of reordering of the way to take summation over $X_r$ and the others.
In order to evaluate this bound, we define
\begin{equation}\label{EqS:Sr_def}
    S(r) \equiv \max_{X_r \subset \Lambda: |X_r| \leq k} \sum_{\substack{X_0,\cdots,X_{r-1} \\ (X_0,\cdots,X_{r-1}) \in \mcl{G}_{r-1}, \\ Y_r \cap X_r \neq \phi}} \prod_{q'=0}^{r-1} \norm{h_{X_{q'}}}.
\end{equation}
for $r \geq 1$. From the locality and extensiveness, we have $S(1) = \max_{X_1} \sum_{X_0: X_0 \cap X_1 \neq \phi} \norm{h_{X_0}} \leq kg$ for example.
We set $S(0)=1$ for the convenience in the inequality below.
Since $Y_r \cap X_r \neq \phi$ implies that some of $X_0,\cdots,X_{r-1}$ should have nonempty intersection with $X_r$, we have
\begin{eqnarray}
    S(r) &\leq& \max_{X_r: |X_r| \leq k} \sum_{r'=0}^{r-1} \sum_{\substack{X_0,\cdots,X_{r-1} \\ (X_0,\cdots,X_{r-1}) \in \mcl{G}_{r-1}, \\ X_{r'} \cap X_r \neq \phi}} \prod_{q'=0}^{r-1} \norm{h_{X_{q'}}} \nonumber \\
    &=& \max_{X_r: |X_r| \leq k} \sum_{r'=0}^{r-1} \sum_{X_{r'}:X_{r'}\cap X_r \neq \phi} \norm{h_{X_{r'}}}\sum_{\substack{X_0,\cdots,X_{r'-1} \\ (X_0,\cdots,X_{r'-1}) \in \mcl{G}_{r'-1}, \\ Y_{r'} \cap X_{r'} \neq \phi}} \prod_{q'=0}^{r-1} \norm{h_{X_{q'}}} \left( \prod_{q'=r'+1}^r \sum_{X_{q'}:Y_{q'}\cap X_{q'}\neq \phi} \norm{h_{X_{q'}}}\right) \nonumber \\
    &\leq& \sum_{r'=0}^{r-1} kg S(r') \frac{(r-1)!}{r'!} (kg)^{r-r'-1},
\end{eqnarray}
for $r \geq 1$.
We use Eqs. (\ref{EqS:Yq_extensive}) and (\ref{EqS:Sr_def}) in order to obtain the last inequality.
The value $S(r)$ satisfies the recursive relation, $S(r)/\{r!(kg)^r\} \leq \frac1r \sum_{r'=0}^{r-1} S(r')/\{r'!(kg)^{r'}\}$, with $S(0)=1$ and it is easy to check that this implies the inequality,
\begin{equation}
    S(r) \leq r! (kg)^r, \quad r = 0,1,\cdots,
\end{equation}
by induction.
Thus, using Eq. (\ref{EqS:Sr_before}) and the low-energy condition $\sum_{X_r} \braket{\psi|h_{X_r}|\psi} \leq \Delta$, we arrive at the upper bound of Eq. (\ref{EqS:Bound_com_inner0}),
\begin{equation}
    \sum_{\substack{X_0,\cdots,X_q \\ (X_0,\cdots,X_q) \in \mcl{G}_q}} \braket{\psi|h_{X_r}|\psi} \prod_{q'\neq r} \norm{h_{X_{q'}}} \leq q! (kg)^q \Delta.
\end{equation}

The other term in Eq. (\ref{EqS:Bound_com_exp_sqrt}) shares the same upper bound by the same calculation.
Summarizing the relations, Eqs. (\ref{EqS:Com_expectation}), (\ref{EqS:Bound_com_exp_prod}), and (\ref{EqS:Bound_com_exp_sqrt}), we finally obtain
\begin{eqnarray}
    \norm{\Pi [H_{q},[H_{q-1},\cdots,[H_1,H_0]]] \Pi} \leq 2^q \sqrt{q!(kg)^q \Delta}\sqrt{q!(kg)^q \Delta} = q! (2kg)^q \Delta. \quad \square
\end{eqnarray}

Compared to the previous results, this theorem accurately captures the locality of Hamiltonians.
Ref. \cite{Hejazi2024-xs-low-energy} has shown the upper bound by
\begin{equation}
    \norm{\Pi_{\leq \Delta} [H_q,[H_{q-1},\cdots,[H_1,H_0]]] \Pi_{\leq \Delta}} \leq 2^q (\Delta')^{q+1} + q! (kg)^q Ng e^{-\order{\frac{\Delta'-\Delta}{g}}},
\end{equation}
where the second term can be negligible by properly defining $\Delta' = \Delta + \order{g \log N}$.
However, this bound essentially corresponds to the $1$-norm scaling $\norm{[H_q,[H_{q-1},\cdots,[H_1,H_0]]]} \in (\sum_{q'=0}^q \norm{H_{q'}})^{q+1} \subset \order{(Ng)^{q+1}}$, and fails to capture many vanishing terms due to the locality in the nested commutator.
On the other hand, as we have computed in Eq. (\ref{EqS:Bound_com_exp_prod}), Theorem \ref{ThmS:nested_com_low_energy} captures this property, which results in the linear scaling in $\Delta$.
It is essential for concluding that Trotterization is advantageous in the low-energy regime $\Delta \in \order{Ng}$ rather than the one in Refs. \cite{Sahinoglu2021-ng-low-energy,Hejazi2024-xs-low-energy}, $\Delta \in \order{N^{1/(p+1)}}$.

Theorem \ref{ThmS:nested_com_low_energy} will be used for proving the Trotter error for $\Gamma \in \order{1}$ in Theorem \ref{Thm:Error_low_energy}.
The one for $\Gamma \notin \order{1}$ relies on the following corollary, which immediately follows from the derivation of Theorem \ref{ThmS:nested_com_low_energy}.

\begin{corollary}\label{CorS:nested_com_low_energy}
\textbf{}

Let $H = \sum_{X \subset \Lambda} h_X$ be a Hamiltonian composed of positive-semidefinite terms $h_X$, and define the projection to the low-energy space $\Pi_{\leq \Delta}$ by Eq. (\ref{Eq:proj_low_energy}).  
For any state $\ket{\psi}$ such that $\braket{\psi|\psi} = 1$ and $\Pi_{\leq \Delta} \ket{\psi} = \ket{\psi}$, the expectation value of the nested commutator is bounded by
\begin{equation}\label{EqS:nested_com_corollary}
    \sum_{X_0,X_1,\cdots,X_q} |\braket{\psi|[h_{X_q},[h_{X_{q-1}},\cdots,[h_{X_1},h_{X_0}]]]|\psi}| \leq q!(2kg)^q \Delta.
\end{equation}
\end{corollary}

\textbf{Proof.---}
In the left hand side of Eq. (\ref{EqS:nested_com_corollary}), the terms with $(X_0,X_1,\cdots,X_q) \in  \mcl{G}_q$ survives, and we have
\begin{equation}
    \sum_{X_0,X_1,\cdots,X_q} |\braket{\psi|[h_{X_q},[h_{X_{q-1}},\cdots,[h_{X_1},h_{X_0}]]]|\psi}| \leq 2^q \max_\sigma \sum_{\substack{X_0,\cdots,X_q \\ (X_0,X_1,\cdots,X_q) \in \mcl{G}_q}} |\braket{\psi|h_{X_q}h_{X_{\sigma(q-1)}}\cdots h_{X_{\sigma(0)}}|\psi}|.
\end{equation}
Although the right hand side differs from Eq. (\ref{EqS:Bound_com_exp_prod}) by the presence of the superscripts $\{\gamma\}$, the same calculation applies to the above formula owing to $\sum_X \braket{\psi|h_X|\psi} \leq \Delta$. $\quad \square$

\section{Trotter errors in the low-energy subspace}\label{SecS:trotter_error}

Here, we prove Theorem \ref{Thm:Error_low_energy}, which gives the Trotter errors for low-energy initial states.
First, we provide the proof for generic cases where the number of partition $\Gamma$ is not specified by Theorem \ref{ThmS:Trot_error_Gamma_notin_1}.
It corresponds to the statement for $\Gamma \notin \order{1}$ in Theorem \ref{Thm:Error_low_energy}, since it is indeed useful for the case with $\Gamma \notin \order{1}$.
After that, we slightly improve the calculation in the derivation of Theorem \ref{ThmS:Trot_error_Gamma_notin_1} for $\Gamma \in \order{1}$, and obtain Corollary \ref{CorS:Trot_error_Gamma_1} as the statement for $\Gamma \in \order{1}$ in Theorem \ref{Thm:Error_low_energy}.
We begin with proving the following theorem especially for $\Gamma \notin \order{1}$.

\begin{theorem}\label{ThmS:Trot_error_Gamma_notin_1}
\textbf{(Trotter errors in the low-energy subspace for $\Gamma \notin \order{1}$)}

Suppose that the $k$-local and $g$-extensive Hamiltonian $H = \sum_X h_X$ ($h_X \geq 0$) is decomposed into $H=\sum_\gamma H_\gamma$, where each $H_\gamma$ is written by $H_\gamma = \sum_X h_X^\gamma$ ($h_X^\gamma \geq 0$).
We consider the error of the $p$-th order Trotter formula $T_p(t)$ within the low-energy subspace having the energy smaller than $\Delta$.
For an arbitrary real number $\epsilon \in (0,1)$, we set two values $p_0$ and $\Delta'$ by
\begin{equation}\label{EqS:p_0_Delta'}
    p_0 = \lceil \log (2k^{-1}N/\epsilon) + 1 \rceil \in \order{\log (N/\epsilon)}, \quad \Delta' = \Delta + 4kg \log \left( \frac{e^2 N}{\epsilon}\right) \in \Delta + \order{g \log(N/\epsilon)}.
\end{equation}
Then, if the time $t$ satisfies
\begin{equation}\label{EqS:time_condition}
    |t| \leq \frac12 (2c_p p_0 kg)^{-1} \in \order{\{\log(N/\epsilon)\}^{-1}},
\end{equation}
with $c_p$ being the number of repetitions in the Trotter formula [See Eq. (\ref{Eq:product_formula})], the error of interest is bounded by
\begin{equation}\label{EqS:error_low_energy_Gamma_generic}
    \varepsilon_{p,\Delta}(t) \equiv \norm{(e^{-iHt}-T_p(t))\Pi_{\leq \Delta}} \leq \frac{4c_p}{p+1} (2c_p p_0 kgt)^p \Delta' t + \epsilon.
\end{equation}
Namely, the error in the low-energy space scales as $\varepsilon_{p,\Delta}(t) \in \order{\{gt \log (N/\epsilon)\}^p \{\Delta + g \log (N/\epsilon)\}t + \epsilon}$.
\end{theorem}

\textbf{Proof.---}
We use the following formula of the Trotter error \cite{childs2021-trotter},
\begin{eqnarray}
    e^{-iHt} - T_p(t) &=& - e^{-iHt} \int_0^t \dd \tau \dv{\tau} \left( e^{iH\tau} T_p(\tau) \right) \nonumber \\
    &=& -i \int_0^t \dd \tau e^{-iH(t-\tau)} T_p(\tau) D_p(\tau), \\
    D_p(\tau) &\equiv& T_p(\tau)^\dagger H T_p(\tau) - i T_p(\tau)^\dagger \dv{\tau} T_p(\tau). \label{EqS:error_bound_Dp}
\end{eqnarray}
Then, the Trotter error in the low-energy subspace is bounded by
\begin{eqnarray}
    \varepsilon_{p,\Delta}(t) &=& \norm{\int_0^t \dd \tau e^{-iH(t-\tau)} T_p(\tau) D_p(\tau) \Pi_{\leq \Delta}} \nonumber \\
    &\leq& \int_0^t \dd \tau \norm{D_p(\tau)\Pi_{\leq \Delta}}.
\end{eqnarray}
The operator $D_p(\tau)$ determines the bound of the Trotter error.
Using the definition of the Trotter formula $T_p(t)$ by Eq. (\ref{Eq:product_formula}), it is expressed by
\begin{equation}\label{EqS:Dp_tau}
    D_p(\tau) = \sum_{\gamma=1}^\Gamma \prod_{v'=1,\cdots,c_p\Gamma}^\leftarrow e^{i \alpha_{v'}t \ad_{H_{\gamma_{v'}}}} H_\gamma -  \sum_{v=1}^{c_p\Gamma} \alpha_v \prod_{v'<v}^\leftarrow e^{i\alpha_{v'}t \ad_{H_{\gamma_{v'}}} } H_{\gamma_v},
\end{equation}
where the operator $\ad_A B$ for operators $A,B$ denotes their commutator $[A,B]=AB-BA$.
A product of operators with arrows is defined by
\begin{equation}
    \prod_{v'=1,\cdots,V'}^\leftarrow O_{v'} = O_{V'} \cdots O_2 O_1, \quad \prod_{v'=1,\cdots,V'}^\rightarrow O_{v'} = O_1 O_2 \cdots O_{V'}.
\end{equation}
Then, we employ the Taylor's theorem on $e^{\tau \ad_A} B$, given by
\begin{equation}\label{EqS:Exp_Taylor}
    e^{\tau \ad_A} B = \sum_{q=0}^{p_0-1} \frac{(\tau \ad_A)^{q} B}{q !}   + \int_0^\tau \dd \tau_1 \frac{\tau_1^{p_0-1}}{(p_0-1)!} e^{(\tau-\tau_1)\ad_A}  (\ad_{A_1})^{p_0}B.
\end{equation}
Repeating this relation for Eq. (\ref{EqS:Dp_tau}) so that the Taylor's reminder can be in $\order{\tau^{p_0}}$ with a certain order $p_0 > p$, we obtain the expansion of $D_p(\tau)$ by
\begin{equation}\label{EqS:Dp_tau_Taylor}
    D_p(\tau) = \sum_{q=p}^{p_0-1} A_q \tau^q + A_{p_0}(\tau), \quad A_{p_0}(\tau) \in \order{\tau^{p_0}}.
\end{equation}
The order $p_0$ is determined later, and we will see that the choice by Eq. (\ref{EqS:p_0_Delta'}) is appropriate.
The operators $A_q$ and $A_{p_0}(\tau)$ are respectively defined by
\begin{eqnarray}
    A_q &=& i^q \left\{ \sum_{\gamma=1}^\Gamma \sum_{\substack{q_1,\cdots,q_{c_p\Gamma} \geq 0 \\ q_1+\cdots+q_V=q}} \prod_{v' \leq c_p\Gamma}^\leftarrow \frac{(\alpha_{v'} \ad_{H_{\gamma_{v'}}})^{q_{v'}}}{q_{v'}!} H_\gamma  - \sum_{v=1}^{c_p\Gamma} \alpha_v \sum_{\substack{q_1,\cdots,q_{v-1} \geq 0 \\ q_1+\cdots+q_{v-1}=q}} \prod_{v'<v}^\leftarrow \frac{(\alpha_{v'} \ad_{H_{\gamma_{v'}}})^{q_{v'}}}{q_{v'}!} H_{\gamma_v}  \right\}, \label{EqS:A_p} \\
    A_{p_0}(\tau) &=& \sum_{\gamma=1}^\Gamma \sum_{v'=1}^{c_p\Gamma} \sum_{\substack{q_1,\cdots,q_{v'}\geq 0 \\ q_{v'} \neq 0 \\ q_1+\cdots+q_{v'}=p_0}} \int_0^\tau \dd \tau_1 C_{v',c_p\Gamma}^{p_0, \{q\}} (\tau,\tau'; H_\gamma) - \sum_{v=1}^{c_p\Gamma} \sum_{v'<v} \sum_{\substack{q_1,\cdots,q_{v'}\geq 0 \\ q_{v'} \neq 0 \\ q_1+\cdots+q_{v'}=p_0}} \int_0^\tau \dd \tau_1 \alpha_v C_{v',v}^{p_0,\{q\}}(\tau,\tau_1; H_{\gamma_v}), \label{EqS:A_p0_reminder}
\end{eqnarray}
where the operator $C_{v',V'}^{p_0, \{q\}}(\tau,\tau_1; H_\gamma)$ with the label $\{q\}=(q_1,\cdots,q_{v'})$ is given by the following formula \cite{childs2021-trotter}:
\begin{equation}
    C_{v',V'}^{p_0, \{q\}}(\tau,\tau_1; H_\gamma) = \frac{i^{p_0}(\tau-\tau_1)^{q_{v'}-1}\tau^{p_0-q_{v'}}}{(q_{v'}-1)! q_{v'-1}! \cdots q_1 !}\left( \prod_{v''; v'< v'' \leq V'}^\leftarrow e^{i \alpha_{v''} \tau \ad_{H_{\gamma_{v''}}}} \right) e^{i \alpha_v \tau_1 \ad_{H_{\gamma_{v'}}}} \left\{ \prod_{v'' \leq v'}^\leftarrow (\alpha_{v''} \ad_{H_{\gamma_{v''}}})^{q_{v''}} \right\} H_\gamma.
\end{equation}
When considering the Trotter error without energy restriction $\norm{e^{-iHt}-T_p(t)}$ \cite{childs2021-trotter}, the choice $p_0=p$ is sufficient for obtaining the commutator-scaling.
This is because the unitary transformation in the integral of $A_p(\tau)$ is irrelevant to the norm.
By contrast, it causes the leakage from the low-energy subspace by the nonlocal operators $e^{-iH_\gamma t}$ or $T_p(t)$, which is the origin of the overestimation of the the Trotter error in the previous studies \cite{Sahinoglu2021-ng-low-energy,Gong2024-wr-low-energy,Hejazi2024-xs-low-energy}.
This is why we expand $D_p(\tau)$ up to the order $p_0>p$ for bounding the Trotter error in the low-energy subspace.

We next substitute the expansion Eq. (\ref{EqS:Dp_tau_Taylor}) into Eq. (\ref{EqS:error_bound_Dp}), and obtain the following bound on the Trotter error:
\begin{eqnarray}
    \varepsilon_{p,\Delta}(t) &\leq& \int_0^t \dd \tau \norm{A_{p_0}(\tau) \Pi_{\leq \Delta}} + \sum_{q=p}^{p_0-1} \frac{t^{q+1}}{q+1} \norm{A_q \Pi_{\leq \Delta}} 
 \nonumber \\
    &\leq& \int_0^t \dd \tau \norm{A_{p_0}(\tau)} + \sum_{q=p}^{p_0-1} \frac{t^{q+1}}{q+1} \norm{\Pi_{>\Delta'} A_q \Pi_{\leq \Delta}} + \sum_{q=p}^{p_0-1} \frac{t^{q+1}}{q+1} \norm{\Pi_{ \leq \Delta'} A_q \Pi_{\leq \Delta'}}. \label{EqS:error_expansion}
\end{eqnarray}
In the second inequality, we introduce a value $\Delta'$ such that $\Delta' > \Delta$, and use the trivial inequality $\norm{\Pi_{\leq \Delta'} A_q \Pi_{\leq \Delta}} \leq \norm{\Pi_{\leq \Delta'} A_q \Pi_{\leq \Delta'}}$.

To evaluate this bound, we use two properties of local and extensive Hamiltonians.
The first property is about their nested commutator \cite{Kuwahara2016-yn,childs2021-trotter}, 
\begin{eqnarray}
    \alpha_\mr{com}^{q+1} &\equiv& \sum_{\gamma,\gamma_1,\cdots,\gamma_q=1}^\Gamma \norm{[H_{\gamma_q},[H_{\gamma_{q-1}},\cdots, [H_{\gamma_1}, H_{\gamma_0}]]]} \nonumber \\
    &=& \sum_{\gamma_0,\gamma_1,\cdots,\gamma_q=1}^\Gamma \sum_{X_0,X_1,\cdots,X_q} \norm{[h_{X_q}^{\gamma_q},[h_{X_{q-1}}^{\gamma_{q-1}},\cdots, [h_{X_1}^{\gamma_1}, h_{X_0}^{\gamma_0}]]]}.
\end{eqnarray}
Considering that the summation of $h_X^\gamma$ over each $\gamma$ covers the terms $\{h_X\}$ in $H$ without degeneracy, it is bounded by
\begin{equation}\label{EqS:norm_nested_com}
    \alpha_\mr{com}^{q+1} = \sum_{X_0,X_1,\cdots,X_q} \norm{[h_{X_q},[h_{X_{q-1}},\cdots, [h_{X_1}, h_{X_0}]]]} \leq q! (2kg)^q Ng
\end{equation}
as well as Eq. (\ref{EqS:nested_com_norm_general}).
The second property is the exponential suppression of excitations by local operators \cite{Arad2016-ak-excitation}.
Under the $k$-local and $J$-extensive Hamiltonian $H$, an inequality 
\begin{equation}\label{EqS:suppressed_excitation}
    \norm{\Pi_{> \Delta'} O_X \Pi_{\leq \Delta}} \leq \norm{O_X} e^{-\frac{\Delta'-\Delta-3g|X|}{4kg}},
\end{equation}
is satisfied for arbitrary $\Delta, \Delta'$ such that $\Delta'>\Delta$.
It means that excitation from states with energy smaller than $\Delta$ to those with energy larger than $\Delta'$ by the local operator $O_X$ becomes extremely small if the gap $\Delta'-\Delta$ exceeds $3g|X|$, which gives the energy scale manageable by $O_X$.
Using these properties, we evaluate the upper bounds of the three terms in Eq. (\ref{EqS:error_expansion}) below.

\textit{Bound on the Taylor's reminder.---} 
We first evaluate the upper bound of the Taylor's reminder $A_{p_0}(\tau)$ [Eq. (\ref{EqS:A_p0_reminder})], and determine the proper order $p_0$.
The form of Eq. (\ref{EqS:A_p0_reminder}) is the same as the one for evaluating the error bound of the $p_0$-th order Trotterization without energy limitation \cite{childs2021-trotter}.
The bound on $C_{v',V'}^{p_0,\{q\}}$ is given by
\begin{equation}
    \norm{C_{v',V'}^{p_0,\{q\}}(\tau,\tau_1; H_\gamma)} = \frac{(\tau-\tau_1)^{q_{v'}-1} \tau^{p_0-q_{v'}}}{(q_{v'}-1)! q_{v'-1}! \cdots q_1 !} \norm{\left\{ \prod_{v'' \leq v'}^\leftarrow ( \ad_{H_{\gamma_{v''}}})^{q_{v''}} \right\} H_\gamma}
\end{equation}
for every $\tau_1 \in [0,\tau]$.
Integrating this over $\tau_1$ and $\tau$, we obtain
\begin{eqnarray}
    \int_0^t \dd \tau \norm{A_{p_0}(\tau)}&\leq& t^{p_0+1} \sum_{\gamma=1}^\Gamma \sum_{v'=1}^{c_p\Gamma} \sum_{\substack{q_1,\cdots,q_{v'}\geq 0 \\ q_{v'} \neq 0 \\ q_1+\cdots+q_{v'}=p_0}} \frac{t^{p_0+1}}{q_{v'}! \cdots q_1!} \norm{\left\{ \prod_{v'' \leq v'}^\leftarrow ( \ad_{H_{\gamma_{v''}}})^{q_{v''}} \right\} H_\gamma} \nonumber \\
    && \qquad \qquad + \sum_{v=1}^{c_p\Gamma} \sum_{v'<v} \sum_{\substack{q_1,\cdots,q_{v'}\geq 0 \\ q_{v'} \neq 0 \\ q_1+\cdots+q_{v'}=p_0}} \frac{t^{p_0+1}}{q_{v'}! \cdots q_1!} \norm{\left\{ \prod_{v'' \leq v'}^\leftarrow ( \ad_{H_{\gamma_{v''}}})^{q_{v''}} \right\} H_{\gamma_v}} \nonumber \\
    &\leq& 2t^{p_0+1} \sum_{v=1}^{c_p\Gamma}  \sum_{\substack{q_1,\cdots,q_{c_p\Gamma} \geq 0 \\ q_1+\cdots+q_{c_p\Gamma}=p_0}} \norm{\left\{ \prod_{v' \leq c_p\Gamma}^\leftarrow ( \ad_{H_{\gamma_{v'}}})^{q_{v'}} \right\} H_{\gamma_v}}. \label{EqS:A_p0_tau_evaluate}
\end{eqnarray}
Since the summation over $v$ or $q_1,\cdots,q_{c_p \Gamma}$ repeats each term $H_\gamma$ for $c_p$ times, it is further bounded by
\begin{eqnarray}
    \int_0^t \norm{A_{p_0}(\tau)}\dd \tau &\leq& 2 (c_p t)^{p_0+1} \sum_{\gamma,\gamma_1,\cdots,\gamma_{p_0}=1}^\Gamma \norm{[H_{\gamma_{p_0}},[H_{\gamma_{p_0-1}},\cdots,[H_{\gamma_1},H_\gamma]]]} \nonumber \\
    &\leq& 2(c_p t)^{p_0+1} p_0 ! (2kg)^{p_0} Ng
    \nonumber \\
    &\leq& e^2 (2 e^{-1} c_p p_0 kgt)^{p_0+1} k^{-1}N.
\end{eqnarray}
We use Eq. (\ref{EqS:norm_nested_com}) for deriving the second inequality, and use the inequality from the Stirling formula, $q! \leq e q^{q+1} e^{-q}$ for the last inequality.
Now, we set the integer $p_0 \in \order{\log(N/\epsilon)}$ by Eq. (\ref{EqS:p_0_Delta'}) and impose the assumption on the time $t$ by Eq. (\ref{EqS:time_condition}).
Under these conditions, we confirm that the above value is bounded by
\begin{equation}\label{EqS:error_expansion_1}
    \int_0^t \norm{A_{p_0}(\tau)}\dd \tau \leq e^{-p_0+1} k^{-1} N \leq \frac{\epsilon}2,
\end{equation}
which gives the half of the negligible part of the Trotter error in Eq. (\ref{EqS:error_low_energy_Gamma_generic}).

\textit{Bound on $\norm{\Pi_{> \Delta'} A_q \Pi_{\leq \Delta}}$.---}
We next evaluate the second term in Eq. (\ref{EqS:error_expansion}) based on the representation of $A_q$ by Eq. (\ref{EqS:A_p}).
In a similar manner to the derivation of Eq. (\ref{EqS:A_p0_tau_evaluate}), we obtain
\begin{eqnarray}
    \norm{\Pi_{> \Delta'} A_q \Pi_{\leq \Delta}} &\leq& 2 \sum_{v=1}^{c_p\Gamma} \sum_{\substack{q_1,\cdots,q_{c_p\Gamma} \geq 0 \\ q_1+\cdots+q_{c_p\Gamma}=q}} \norm{ \Pi_{>\Delta'} \prod_{v' \leq c_p\Gamma}^\leftarrow ( \ad_{H_{\gamma_{v'}}})^{q_{v'}} H_{\gamma_v} \Pi_{\leq \Delta}} \nonumber \\
    &\leq& 2 (c_p)^{q+1} \sum_{X_0,X_1,\cdots,X_q} \norm{\Pi_{>\Delta'}[h_{X_p},[h_{X_{p-1}},\cdots,[h_{X_1},h_{X_0}]]] \Pi_{\leq \Delta}}. \label{EqS:leakage_evaluate}
\end{eqnarray}
Then, we can use Eq. (\ref{EqS:suppressed_excitation}), which gives the suppressed excitation by local operators, for the nested commutator $[h_{X_p},[h_{X_{p-1}},\cdots,[h_{X_1},h_{X_0}]]]$ since it acts on at most $(q+1)k$ sites.
Also using the inequality, Eq. (\ref{EqS:norm_nested_com}), we arrive at
\begin{eqnarray}
    \norm{\Pi_{> \Delta'} A_q \Pi_{\leq \Delta}} &\leq& 2 (c_p)^{q+1} \sum_{X_0,X_1,\cdots,X_q} \norm{[h_{X_q},[h_{X_{q-1}},\cdots,[h_{X_1},h_{X_0}]]]} e^{-\frac{\Delta'-\Delta-3g (q+1)k}{4kg}} \nonumber \\
    &\leq& 2(c_p)^{q+1} q! (2kg)^q Ng e^{3(q+1)/4} e^{-\frac{\Delta'-\Delta}{4kg}} \nonumber \\
    &\leq& e^2 (2 e^{-1/4} c_p qkg)^{q+1} k^{-1} N e^{-\frac{\Delta'-\Delta}{4kg}},  
\end{eqnarray}
where we use $q! \leq e q^{q+1} e^{-q}$ from the Stirling formula.
Therefore, we have
\begin{eqnarray}
    \sum_{q=p}^{p_0-1} \frac{t^{q+1}}{q+1} \norm{\Pi_{> \Delta'} A_q \Pi_{\leq \Delta}} &\leq& e^2 \sum_{q=p}^\infty (2e^{-1/4}c_p p_0 kg)^{q+1} k^{-1} N e^{-\frac{\Delta'-\Delta}{4kg}} \nonumber \\
    &\leq& e^2 2^{-p} k^{-1} N e^{-\frac{\Delta'-\Delta}{4kg}} \nonumber \\
    &\leq& \frac{\epsilon}2. \label{EqS:error_expansion_2}
\end{eqnarray}
The final inequality comes from the choice of $\Delta'$ by Eq. (\ref{EqS:p_0_Delta'}).

\textit{Bound on $\norm{\Pi_{\leq \Delta'} A_q \Pi_{\leq \Delta'}}$.---}
Finally, we evaluate the third term in Eq. (\ref{EqS:error_expansion}).
Since the operator $A_q$ defined by Eq. (\ref{EqS:A_p}) is hermitian, its norm in the low-energy subspace is represented by
\begin{eqnarray}
  \norm{\Pi_{\leq \Delta'} A_q \Pi_{\leq \Delta'}} &=& \max_{\ket{\psi}; \braket{\psi|\psi}=1} |\braket{\psi|\Pi_{\leq \Delta'} A_q \Pi_{\leq \Delta'}|\psi}| \nonumber \\
  &=& \max_{\substack{\ket{\psi}; \braket{\psi|\psi}=1 \\ \Pi_{\leq \Delta'}\ket{\psi} = \ket{\psi}}} |\braket{\psi|A_q|\psi}|
\end{eqnarray}
In the calculation below, we suppose that the low-energy state $\ket{\psi}$ satisfies $\braket{\psi|\psi}=1$ and $\Pi_{\leq \Delta'}\ket{\psi} = \ket{\psi}$.
In a similar way to Eqs. (\ref{EqS:A_p0_tau_evaluate}) and (\ref{EqS:leakage_evaluate}), we obtain the bound,
\begin{eqnarray}
    |\braket{\psi|A_q|\psi}| &\leq& 2 \sum_{v=1}^{c_p\Gamma} \sum_{\substack{q_1,\cdots,q_{c_p\Gamma} \geq 0 \\ q_1+\cdots+q_{c_p\Gamma}=q}} \left| \bra{\psi} \prod_{v' \leq c_p\Gamma}^\leftarrow ( \ad_{H_{\gamma_{v'}}})^{q_{v'}} H_{\gamma_v} \ket{\psi} \right| \nonumber \\
    &\leq& 2 (c_p)^{q+1} \sum_{X_0,X_1,\cdots,X_p} |\braket{\psi|[h_{X_p},[h_{X_{p-1}},\cdots,[h_{X_1},h_{X_0}]]]|\psi}|. \label{EqS:A_q_low_evaluate}
\end{eqnarray}
The right hand side includes the nested commutator in the low-energy subspace, which can be evaluated by Corollary \ref{CorS:nested_com_low_energy}.
As a result, the third term of Eq. (\ref{EqS:error_expansion}) is bounded by
\begin{eqnarray}
    \sum_{q=p}^{p_0-1} \frac{t^{q+1}}{q+1} \norm{\Pi_{\leq \Delta'} A_q \Pi_{\leq \Delta'}} &\leq& \frac{2c_p}{p+1} \sum_{q=p}^{p_0-1}q!(2 c_p kg t)^q \Delta't  \nonumber \\
    &\leq& \frac{2c_p}{p+1} \Delta't \sum_{q=p}^\infty (2c_p p_0kgt)^q = \frac{4c_p}{p+1} (2c_p p_0 kgt)^p \Delta't. \label{EqS:error_expansion_3}
\end{eqnarray}
In the second inequality, we use the condition on the time $t$ by Eq. (\ref{EqS:time_condition}).

\textit{Error bound in total.---} 
The error $\varepsilon_{p,\Delta}(t)$ is bounded by Eq. (\ref{EqS:error_expansion}).
Summarizing the results in Eqs. (\ref{EqS:error_expansion_1}), (\ref{EqS:error_expansion_2}), and (\ref{EqS:error_expansion_3}), we conclude Eq. (\ref{EqS:error_low_energy_Gamma_generic}) as its upper bound. $\quad \square$

We next focus on the case $\Gamma \in \order{1}$.
The statement of Theorem \ref{Thm:Error_low_energy} for $\Gamma \in \order{1}$ is given by the following corollary.
\begin{corollary}\label{CorS:Trot_error_Gamma_1}
\textbf{(Trotter errors in the low-energy subspace for $\Gamma \in \order{1}$)}

Consider a Hamiltonian $H = \sum_X h_X$, which is the same as Theorem \ref{ThmS:Trot_error_Gamma_notin_1}.
For an arbitrary real number $\epsilon \in (0,1)$, we set a value $\Delta'$ by
\begin{equation}\label{EqS:Delta'_Gamma_1}
    \Delta' = \Delta + 4kg \log (2^{-p+1} k^{-1}N/\epsilon) \in \Delta + \order{g \log (N/\epsilon)}.
\end{equation}
Then, if the time $t$ satisfies
\begin{equation}\label{EqS:time_condition_Gamma_1}
    |t| \leq \frac1e (2c_p \Gamma kg)^{-1},
\end{equation}
the Trotter error in the low-energy subspace is bounded by
\begin{equation}\label{EqS:error_low_energy_Gamma_1}
    \varepsilon_{p,\Delta} (t) \equiv \norm{(e^{-iHt}-T_p(t)) \Pi_{\leq \Delta}} \leq \frac{2c_p\Gamma}{(1-e^{-1})(p+1)} (2c_p\Gamma kgt)^p \Delta' t + \epsilon.
\end{equation}
\end{corollary}

\textbf{Proof.---}
The proof is almost parallel to the one for Theorem \ref{ThmS:Trot_error_Gamma_notin_1}.
We start from Eq. (\ref{EqS:error_expansion}) with using $p_0$ and $\Delta'$ different from Eq. (\ref{EqS:p_0_Delta'}), which are determined later.
We modify the way of evaluating the summation over $q_1,\cdots, q_{c_p\Gamma}$ in Eqs. (\ref{EqS:A_p0_tau_evaluate}), (\ref{EqS:leakage_evaluate}), and (\ref{EqS:A_q_low_evaluate}).
We first focus on the first term of Eq. (\ref{EqS:error_expansion}),
\begin{equation}
    \int_0^t \norm{A_{p_0}(\tau)}\dd \tau \leq 2t^{p_0+1} \sum_{v=1}^{c_p\Gamma}  \sum_{\substack{q_1,\cdots,q_{c_p\Gamma} \geq 0 \\ q_1+\cdots+q_{c_p\Gamma}=\tilde{p}_0}} \frac{1}{q_1! \cdots q_{c_p\Gamma} !} \norm{\left\{ \prod_{v' \leq c_p\Gamma}^\leftarrow ( \ad_{H_{\gamma_{v'}}})^{q_{v'}} \right\} H_{\gamma_v}},
\end{equation}
from Eq. (\ref{EqS:A_p}).
Since each nested commutator in the right hand side has a form $[H_{\gamma_{p_0}},\cdots, [H_{\gamma_1},H_{\gamma_0}]]$, we apply Eq. (\ref{EqS:norm_nested_com}) for evaluating its norm.
This results in
\begin{eqnarray}
    \int_0^t \dd \tau \norm{A_{p_0}(\tau)} &\leq& 2 \sum_{v=1}^{c_p\Gamma}  \sum_{\substack{q_1,\cdots,q_{c_p \Gamma} \geq 0 \\ q_1+\cdots+q_{c_p\Gamma}=p_0}}  \frac{p_0!}{q_1! \cdots q_{c_p\Gamma} !}(2kg t)^{p_0} Ng t\nonumber \\
    &=& (2c_p\Gamma kgt)^{p_0+1} k^{-1} N. \label{EqS:A_p0_Gamma_1}
\end{eqnarray}
For small time $t$ satisfying Eq. (\ref{EqS:time_condition_Gamma_1}), we set $p_0$ by
\begin{equation}
    p_0 = \lceil \log (2k^{-1}N/\epsilon) \rceil -1 
\end{equation}
instead of Eq. (\ref{EqS:p_0_Delta'}).
Then, Eq. (\ref{EqS:A_p0_Gamma_1}) becomes smaller than $\epsilon/2$.

Similarly, the second term of Eq. (\ref{EqS:error_expansion}) is bounded by
\begin{eqnarray}
    \sum_{q=p}^{p_0-1} \frac{t^{q+1}}{q+1} \norm{\Pi_{> \Delta'} A_q \Pi_{\leq \Delta}} &\leq& \sum_{q=p}^{p_0-1} \frac{2t^{q+1}}{q+1} \sum_{v=1}^{c_p\Gamma} \sum_{\substack{q_1,\cdots,q_{c_p\Gamma} \geq 0 \\ q_1+\cdots+q_{c_p\Gamma}=q}} \frac{1}{q_1 ! \cdots q_{c_p\Gamma} !}\norm{ \Pi_{>\Delta'} \prod_{v' \leq c_p\Gamma}^\leftarrow ( \ad_{H_{\gamma_{v'}}})^{q_{v'}} H_{\gamma_v} \Pi_{\leq \Delta}} \nonumber \\
    &\leq& c_p\Gamma \sum_{q=p}^{p_0-1} \sum_{\substack{q_1,\cdots,q_{c_p\Gamma} \geq 0 \\ q_1+\cdots+q_{c_p\Gamma}=q}} \frac{q!}{q_1 ! \cdots q_{c_p\Gamma} !} (2kgt)^q Ngt e^{-\frac{\Delta'-\Delta - 3 (q+1)kg}{4kg}} \nonumber \\
    &\leq& \frac12 \sum_{q=p}^{p_0-1} (2 e^{3/4} c_p\Gamma kgt)^{p+1} k^{-1} N e^{-\frac{\Delta'-\Delta}{4kg}} \leq 2^{-p} k^{-1} N e^{-\frac{\Delta'-\Delta}{4kg}}. \label{EqS:error_expansion_2_Gamma_1}
\end{eqnarray}
Setting $\Delta'$ by Eq. (\ref{EqS:Delta'_Gamma_1}) ensures that this upper bound becomes smaller than $\epsilon/2$.
Finally, in a similar manner, the last term of Eq. (\ref{EqS:error_expansion}) is bounded by
\begin{eqnarray}
    \sum_{q=p}^{p_0-1} \frac{t^{q+1}}{q+1} \norm{\Pi_{\leq \Delta'} A_q \Pi_{\leq \Delta'}} &\leq&  \sum_{q=p}^{p_0-1} \frac{2t^{q+1}}{q+1} \max_{\substack{\ket{\psi} \\ \Pi_{\leq \Delta'} \ket{\psi}= \ket{\psi}}}\sum_{v=1}^{c_p\Gamma} \sum_{\substack{q_1,\cdots,q_{c_p\Gamma} \geq 0 \\ q_1+\cdots+q_{c_p\Gamma}=q}} \frac1{q_1! \cdots q_{c_p\Gamma} !}\left| \bra{\psi} \prod_{v' \leq c_p\Gamma}^\leftarrow ( \ad_{H_{\gamma_{v'}}})^{q_{v'}} H_{\gamma_v} \ket{\psi} \right| \nonumber \\
    &\leq& c_p\Gamma \sum_{q=p}^{p_0-1} \frac2{q+1} \sum_{\substack{q_1,\cdots,q_{c_p\Gamma} \geq 0 \\ q_1+\cdots+q_{c_p\Gamma}=q}} \frac{q!}{q_1 ! \cdots q_{c_p\Gamma} !} (2kgt)^q \Delta't \nonumber \\
    &\leq& \frac{2c_p\Gamma}{p+1} \sum_{q=p}^\infty (2c_p\Gamma kgt)^q \Delta' t \leq \frac{2c_p\Gamma}{(p+1)(1-e^{-1})} (2c_p\Gamma kgt)^p \Delta' t. \label{EqS:error_expansion_3_Gamma_1}
\end{eqnarray}
We use Corollary \ref{CorS:nested_com_low_energy} for the second inequality and the assumption, Eq. (\ref{EqS:time_condition_Gamma_1}), for the last inequality.
Summarizing the results of Eqs. (\ref{EqS:A_p0_Gamma_1}), (\ref{EqS:error_expansion_2_Gamma_1}), and (\ref{EqS:error_expansion_3_Gamma_1}), we complete the proof of Eq. (\ref{EqS:error_low_energy_Gamma_1}). $\quad \square$

\section{Exact scaling of the Trotter number}\label{SecS:trotter_number}

Here, we derive the exact scaling of the Trotter step $r$, required for simulating low-energy states.
As discussed in the main text, we have to determine $r$ so that
\begin{equation}\label{EqS:algorithm_error_requirement}
    \norm{\left\{ e^{-iHt} - ( T_p(t/r) )^r \right\} \Pi_{\leq \Delta}} \leq \varepsilon.
\end{equation}
can be satisfied.
Then, the Trotter number $r$ can be determined as follows.

\begin{proposition}
\textbf{(Exact scaling of Trotter number $r$)}

Suppose that the time renormalized by the extensiveness, $gt$, and the inverse error $1/\varepsilon$ are at most $\poly{N}$.
Then, there exists a value $r$ satisfying
\begin{equation}\label{EqS:exact_trotter_number}
    r \in \begin{cases}
        \order{gt \left( \frac{\Delta t + gt \log (N/\varepsilon)}{\varepsilon} \right)^\frac1p } & (\text{if $\Gamma \in \order{1}$}) \\
        \order{gt \log (N/\epsilon) \left( \frac{\Delta t + gt \log (N/\varepsilon)}{\varepsilon} \right)^\frac1p} & (\text{otherwise}),
    \end{cases}
\end{equation}
which qualifies as the Trotter number satisfying Eq. (\ref{EqS:algorithm_error_requirement}).

\end{proposition}

\textbf{Proof.---} 
We first consider the case with $\Gamma \in \order{1}$.
We set $\epsilon = \varepsilon / (2r)$ and use Corollary \ref{CorS:Trot_error_Gamma_1}, while we will confirm the condition Eq. (\ref{EqS:time_condition_Gamma_1}) for the Trotter step $t/r$ later.
Let us define a value $r_0$ by
\begin{equation}\label{EqS:r0_trotter}
    r_0 = gt \left( \frac{\Delta t + gt \log (N/\varepsilon)}{\varepsilon} \right)^{\frac1p}.
\end{equation}
Then, it is sufficient to show that there exists a constant $A>1$ such that 
\begin{equation}\label{EqS:A_trotter_condition}
    \frac{(gt)^p\{ \Delta t + gt \log (N A r_0 / \varepsilon) \}}{(A r_0)^p} \leq \varepsilon,
\end{equation}
where the Trotter number $r$ is equal to $A r_0$.
The left hand side is calculated as follows:
\begin{eqnarray}
    \frac{(gt)^p\{ \Delta t + gt \log (N A r_0 / \varepsilon) \}}{(A r_0)^p} &=& \frac{(gt)^p \{ \Delta t + gt \log (N/\varepsilon) \}}{(r_0)^p} A^{-p}\frac{\Delta t + gt \log (N/\varepsilon) + gt \log A r_0}{\Delta t + gt \log (N/\varepsilon)} \nonumber \\
    &\leq& \varepsilon A^{-p} \left( 1 + \frac{\log A}{\log (N/\varepsilon)} + \frac{\log r_0}{\log (N/\varepsilon)}\right) \nonumber \\
    &\leq& \varepsilon A^{-p+\frac12} \left( 2 + \frac{\log r_0}{\log (N/\varepsilon)}\right).
\end{eqnarray}
In the last line, we use $\log A \leq A^{1/2}$ for $A>1$ and $1/\log (N/\varepsilon) < 1$ for sufficiently large $N$ or $1/\varepsilon$.
When the time $t$ and the inverse error $1/\varepsilon$ are both in $\poly{N}$, we have
\begin{eqnarray}
    \log r_0 &\leq& \log (gt) + \frac1p \log \left( \frac{Ngt+gt \log (N/\varepsilon)}{\varepsilon} \right) \in \order{\log N},
\end{eqnarray}
where we use $\Delta \leq Ng$ in the first inequality.
Thus, there exists a constant positive value $B$ such that $\log r_0 / \log (N/\varepsilon) \leq B$.
When we set the positive constant $A$ by $A=(2+B)^{2/(2p-1)}$, Eq. (\ref{EqS:A_trotter_condition}) is satisfied.
We finally discuss the satisfaction of Eq. (\ref{EqS:time_condition_Gamma_1}) for the Trotter step $t/r$ to apply Corollary \ref{CorS:Trot_error_Gamma_1}.
According to Eq. (\ref{EqS:r0_trotter}), it scales as
\begin{eqnarray}
    \frac{t}{r} &\in& \order{g^{-1} \left( \frac{\Delta t + gt \log (N/\varepsilon)}{\varepsilon}\right)^{-\frac1p}}.
\end{eqnarray}
On the other hand, Eq. (\ref{EqS:time_condition_Gamma_1}) demands $\frac{t}r \in \order{g^{-1}}$, where the constant factor depends only on $k,\Gamma,p$.
Thus, under sufficiently large $\Delta t$, $gt$, $N$, or $1/\varepsilon$, the condition Eq. (\ref{EqS:time_condition_Gamma_1}) is automatically satisfied.

The other case including $\Gamma \notin \order{1}$ is discussed in a similar manner.
We set a value $r_0$, given by
\begin{equation}
    r_0 = gt \log (N/\varepsilon) \left( \frac{\Delta t + gt \log (N/\varepsilon)}{\varepsilon} \right)^{\frac1p}.
\end{equation}
Then, based on Theorem \ref{ThmS:Trot_error_Gamma_notin_1}, we seek for a positive constant $A$ such that 
\begin{equation}
    \frac{\{gt \log (NAr_0/\varepsilon)\}^p \{ \Delta t + gt \log (N A r_0 / \varepsilon) \}}{(A r_0)^p} \leq \varepsilon.
\end{equation}
We obtain $r = A r_0$ obeying Eq. (\ref{EqS:exact_trotter_number}) in a similar manner. $\quad \square$

\section{Numerical demonstration on frustration-free Hamiltonians}\label{SecS:Numerical}

\begin{figure*}
    \centering
    \includegraphics[width=18cm]{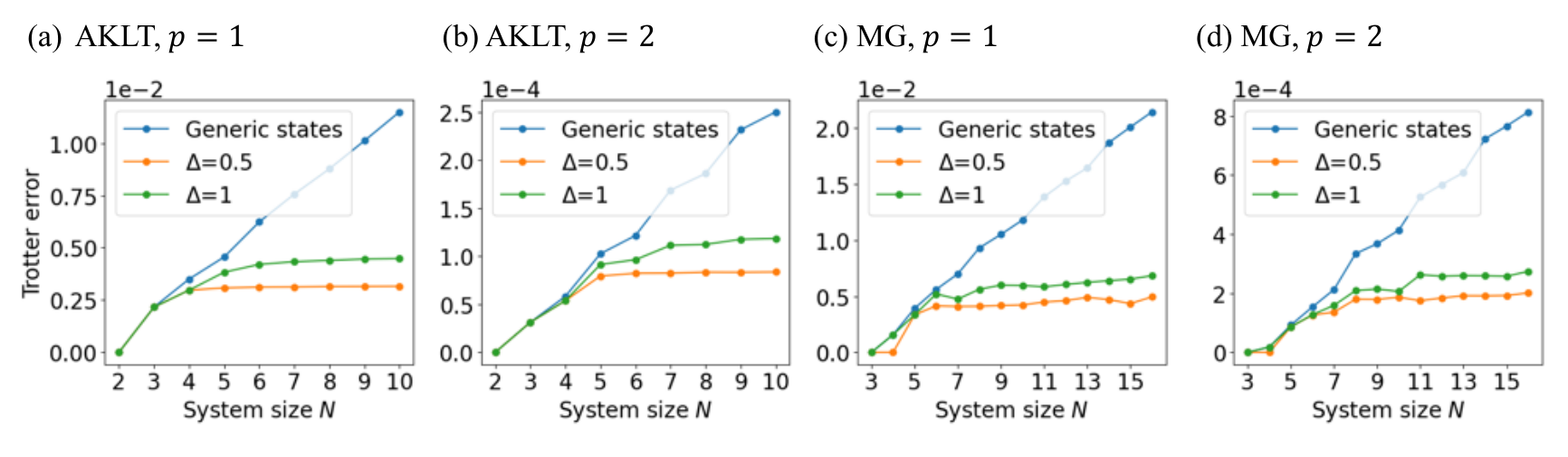}
    \caption{System size dependence of the Trotter error in the AKLT and MG Hamiltonians. The blue lines denote the error for arbitrary initial states, $\norm{e^{-iHt}-T_p(t)}$. The green or orange lines show the errors for initial states in the low-energy subspace, $\norm{(e^{-iHt}-T_p(t)) \Pi_{\leq \Delta}}$. The integer $p$ denotes the Trotter order.}
    \label{Fig:numerical_size_dep}
\end{figure*}

Here, we numerically demonstrate Theorem \ref{Thm:Error_low_energy}, i.e., we will see that the scaling of the Trotter error predicted by it matches the one for some Hamiltonians well.
Theorem \ref{Thm:Error_low_energy} applies to generic local Hamiltonians with shifting every local term by $h_X \to h_X + \norm{h_X}$ for guaranteeing its positive-semidefiniteness.
As discussed in the main text, the error reduction appears in the low-energy regime $\Delta \in o(Ng)$.
Although it is comparably relaxed from the one in the previous studies \cite{Sahinoglu2021-ng-low-energy,Hejazi2024-xs-low-energy}, $\Delta \in o(N^{1/(p+1)}g)$, it still demands the Hamiltonian to be nearly frustration-free.
Thus, we hereby concentrate on the following completely frustration-free Hamiltonians.
The first one is the Affleck-Kennedy-Lieb-Tasaki (AKLT) Hamiltonian,
\begin{equation}
    H_\mr{AKLT} = \sum_{i=1}^{N-1} h_{i,i+1}, \quad h_{i,i+1} = P \left[ (\bmvec{S}_i+\bmvec{S}_{i+1})^2 = 2(2+1) \right],
\end{equation}
where each $\bmvec{S}_i$ is a spin-$1$ operator at a site $i$ \cite{Affleck1988-ia}.
The operator $P[\cdot]$ means a projection to the subspace satisfying the argument.
The AKLT Hamiltonian is a typical example whose gapped ground state has a symmetry protected topological order.
The second example is the Majumdar-Ghosh (MG) Hamiltonian defined by
\begin{equation}
    H_\mr{MG} = \sum_{i=1}^{N-2} h_{i,i+1,i+2}, \quad h_{i,i+1, i+2} = P \left[ (\bmvec{S}_i+\bmvec{S}_{i+1}+\bmvec{S}_{i+2})^2 = \frac32 \left(\frac32+1\right) \right],
\end{equation}
where each $\bmvec{S}_i$ is a spin-$1/2$ operator \cite{Majumdar1969-cn-1,Majumdar1969-rw-2}.
The MG Hamiltonian is known to have two exactly-degenerate ground states.
We demonstrate how the Trotter error $\varepsilon_{p,\Delta}(t)$ depends on the system size $N$ and the energy bound $\Delta$ by exact diagonalization.

\begin{figure*}
    \centering
    \includegraphics[width=18cm]{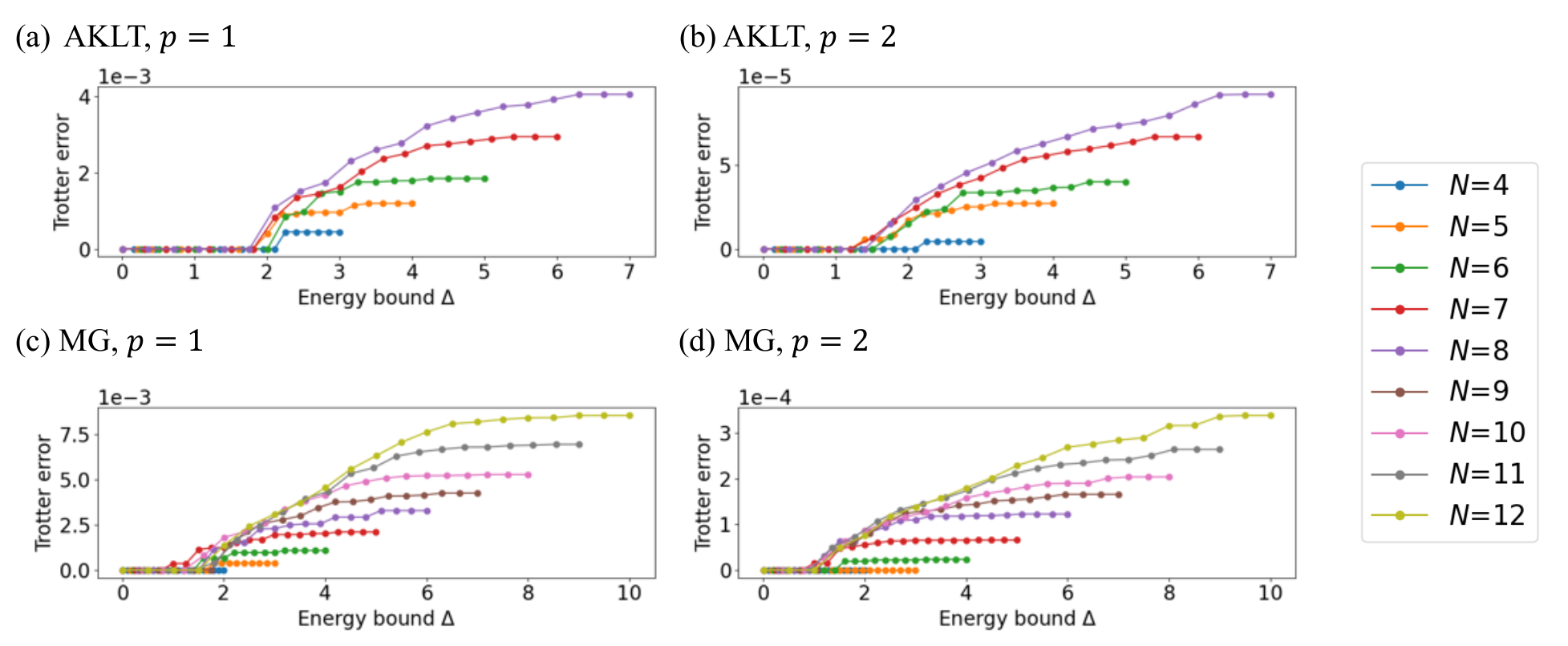}
    \caption{$\Delta$-dependence of the Trotter error, where $\Delta$ denotes the maximal energy scale of the initial state. }
    \label{fig:numerical_Delta_dep}
\end{figure*}

We organize the Trotterization $T_p(t)$ for the orders $p=1,2$ by the Lie-Suzuki-Trotter formula, Eq. (\ref{Eq:Lie-Suzuki-Trotter}).
We decompose the AKLT Hamiltonian by $H_\mr{AKLT} = H_1+H_2$ with $H_1 = \sum_{i;\text{odd}} h_{i,i+1}$ and $H_2 = \sum_{i;\text{even}} h_{i,i+1}$.
Similarly, we decompose the MG Hamiltonian $H_\mr{MG}$ into three terms depending on the site $i$ modulo $3$.
Figure \ref{Fig:numerical_size_dep} shows the system-size dependence of the Trotter error.
We fix $\Delta = 0.5$ or $\Delta = 1.0$ and consider the evolution time $t = 0.1$.
Without any energy restriction, the Trotter errors are approximately proportional to the system size (See the blue solid lines), which is consistent with the commutator-scaling errors (See the first row in  Table \ref{Table:Comparison}).
By contrast, the one for low-energy initial states (the orange and green solid lines) grow much slower.
This behavior matches the logarithmic system-size dependence by Theorem \ref{Thm:Error_low_energy}, rather than the polynomial dependence predicted by the previous studies \cite{Sahinoglu2021-ng-low-energy,Hejazi2024-xs-low-energy}.
On the other hand, Fig. \ref{fig:numerical_Delta_dep} shows the dependence on the energy bound $\Delta$.
We change the value $\Delta$ from $0$ to $\norm{H}$, and thus each rightmost point means the Trotter error for arbitrary initial states.
For the points in the left sides (i.e., the regions with $\Delta \ll \norm{H}$), we observe a slope which upper-bounds the Trotter error.
The slope is approximately linear in the energy bound $\Delta$ and almost independent of the system size $N$.

This result supports that Trotterization becomes exponentially more efficient in the system size for low-energy states than for generic initial states.

\section{Weakly-correlated states with low-energy expectation value}\label{SecS:weakly_correlated}

Throughout the main text, we consider a low-energy initial state $\ket{\psi}$ satisfying $\Pi_{\leq \Delta} \ket{\psi} = \ket{\psi}$, which is completely closed in the low-energy subspace.
As a straightforward extension based on the concentration bound or the Chernoff bound \cite{Anshu2016-gc-concentration,Kuwahara2016-gh-Chernoff}, we see that the reduction of the error and cost of Trotterization can be present also for weakly-correlated initial states having a low-energy expectation value.

We suppose that the initial state $\ket{\psi}$ is a product state whose energy expectation value  $\braket{\psi|H|\psi}$ is sufficiently small.
The Hamiltonian $H$ is assumed to be composed of finite-ranged Hamiltonians.
Then, the concentration bound dictates that the weight of eigenstates highly deviated from $\Delta$ exponentially decays as
\begin{equation}\label{EqS:concentration_bound}
    \braket{\psi|\Pi_{>\braket{\psi|H|\psi} +x}|\psi} \leq \exp \left( - c \frac{x^2}{Ng^2}\right),
\end{equation}
where $c$ is a positive $\order{1}$ constant  determined by the locality $k$, the extensiveness $g$, and the lattice geometry \cite{Anshu2016-gc-concentration}.
The Trotter error for the initial product state $\ket{\psi}$ is bounded by
\begin{eqnarray}
    \norm{\left( e^{-iHt} - \{T_p(t/r)\}^r \right) \ket{\psi}} &\leq& \norm{\left( e^{-iHt} - \{T_p(t/r)\}^r \right) \Pi_{\leq \braket{\psi|H|\psi}+x}} + \norm{e^{-iHt} - \{T_p(t/r)\}^r} \cdot \norm{\Pi_{> \braket{\psi|H|\psi} + x} \ket{\psi}} \nonumber \\
    &\leq& r \varepsilon_{p,\braket{\psi|H|\psi}+x}(t/r) + 2 \norm{\Pi_{> \braket{\psi|H|\psi} + x} \ket{\psi}},
\end{eqnarray}
based on the triangle inequality.
In order to execute Hamiltonian simulation with the guaranteed accuracy $\varepsilon$, it suffices to choose the value $x$ by 
\begin{equation}\label{EqS:concentration_width}
    x = \sqrt{\frac{2N}{c} \log (4/\varepsilon)} g\in \Theta \left( \sqrt{N \log (1/\varepsilon)} g \right),
\end{equation}
which results in $\norm{\Pi_{> \braket{\psi|H|\psi} + x} \ket{\psi}} \leq \varepsilon/4$ by Eq. (\ref{EqS:concentration_bound}), and to determine an integer $r$ satisfying $r \varepsilon_{p,\braket{\psi|H|\psi}+x}(t/r) \leq \varepsilon/2$.
The Trotter number $r$ scales as
\begin{equation}
    r \in \Theta \left( gt \left( \frac{\braket{\psi|H|\psi} t + \sqrt{N\log (1/\varepsilon)}gt}{\varepsilon}\right)^{1/p}\right),
\end{equation}
which is obtained in a similar manner to Supplementary Materials \ref{SecS:trotter_number}.
Therefore, compared to the case for arbitrary initial states \cite{childs2021-trotter}, there exists constant speedup under $\braket{\psi|H|\psi} \leq \mr{Const.} \times Ng$, and the scaling of the Trotter number is improved under $\braket{\psi|H|\psi} \in o(Ng)$.
Under the deeper low-energy regime  $\braket{\psi|H|\psi} \in \order{N^{1/2}g}$, its system-size dependence can be $\order{N^{1/(2p)}}$, which is quadratically better than the one for arbitrary initial states (See the first row in Table \ref{Table:Comparison}).

In contrast to being closed in the low-energy subspace, it is easy to check whether the initial product state has small energy expectation value by classical computation.
Instead, the improvement is at most quadratic in $N$ due to the energy distribution. 
We also note that the advantage for initial states with a low-energy expectation value becomes visible for the first time by our result.
When we discuss with the previous results \cite{Sahinoglu2021-ng-low-energy,Hejazi2024-xs-low-energy}, the value $\braket{\psi|H|\psi} + x$ is demanded to be $o(N^{1/(p+1)})$ for the advantage in the scaling to be present.
However, this cannot be satisfied since the width of the energy distribution $x$ scales as Eq. (\ref{EqS:concentration_width}).
Finally, while we focus on product states under a Hamiltonian with finite-ranged interactions, the concentration bound or the Chernoff bound exist for weakly-correlated states (i.e., states with exponentially-decaying correlation functions or product states with finite-depth local unitary gates) \cite{Anshu2016-gc-concentration} or for generic $k$-local and $g$-extensive Hamiltonians \cite{Kuwahara2016-gh-Chernoff}.
The advantage is present also for these cases, while the width $x$ is modified from Eq. (\ref{EqS:concentration_width}) poly-logarithmically in $1/\varepsilon$.

Recently, it has been proven that the Trotter error can reduce for initial states having large entanglement \cite{Zhao-prl2022-random,zhao2024-entanglement}.
Moreover, it has been shown that it is impossible to improve the Trotter error for generic product states from the worst case error $\order{Nt^{p+1}}$ \cite{childs-prl2019-pf}, since there exists a certain product state having the Trotter error proportional to $N$ \cite{zhao2024-entanglement}.
On the other hand, as opposed to expectation from these studies, our result indicates that weakly-correlated initial states with a small amount of entanglement can generally have a smaller Trotter error if we additionally assume that they have small energy expectation values.

\end{document}